\documentclass[aps,prb,twocolumn,floats,floatfix,epsfig,pdflatex,longbiblography]{revtex4-1}
\usepackage{amssymb,amsbsy,amsmath,mathrsfs}
\usepackage{epsfig,psfrag,subfigure}
\usepackage{graphicx,float}
\usepackage{times}
\usepackage{color}
\usepackage{subfigure}
\usepackage{setspace}
\usepackage{bm} 
\usepackage{xr}
\usepackage{stackengine}
\usepackage{natbib,url,hyperref}
\hypersetup{colorlinks,allcolors=blue}
\usepackage{filecontents}
\usepackage{comment}

\makeatletter
\newcommand*{\addFileDependency}[1]{
  \typeout{(#1)}
  \@addtofilelist{#1}
  \IfFileExists{#1}{}{\typeout{No file #1.}}
}
\makeatother

\newcommand\bea{\begin{eqnarray}}
\newcommand\eea{\end{eqnarray}}
\newcommand\beq{\begin{equation}}
\newcommand\eeq{\end{equation}}

\newcommand{\non}{\nonumber}

\newcommand{\ta}{\theta}
\newcommand{\om}{\omega}

\newcommand{\la}{\langle}
\newcommand{\ra}{\rangle}

\newcommand{\bra}[1]{\langle #1|}
\newcommand{\ket}[1]{|#1\rangle}

\begin{document}
\title{Family-Vicsek dynamical scaling and Kardar-Parisi-Zhang-like superdiffusive growth of surface roughness in a driven one-dimensional quasiperiodic  model}
\author{Sreemayee Aditya$^1$, Nilanjan Roy$^{2,3}$}

\affiliation{$^1$Centre for High Energy Physics, Indian Institute of 
Science, Bengaluru 560012, India \\
$^2$Centre for Condensed Matter Theory, Department of Physics, Indian Institute of 
Science, Bengaluru 560012, India \\
$^3$ Division of Physics and Applied Physics, School of Physical and Mathematical Sciences, Nanyang Technological University, Singapore 637371}
\begin{abstract}
The investigation of the dynamical universality classes of quantum systems is an important, and rather less explored aspect of non-equilibrium physics.
In this work, considering
the out-of-equilibrium dynamics of spinless fermions 
in a one-dimensional quasiperiodic model with and without periodic driving, 
we report the existence of the dynamical one-parameter Family-Vicsek (FV) scaling of the ``quantum surface-roughness" associated with the particle-number fluctuations. 
In the absence of periodic driving, the model is interestingly shown to host a subdiffusive critical phase with anomalous FV scaling exponents,  separated by two subdiffusive critical lines and a triple point from other phases.
Our analysis on the fate of the critical phase in the presence of (inter-phase) driving indicates that the critical phase is quite fragile, and has a tendency to get absorbed into the delocalized or localized regimes depending on the driving parameters, especially in the slow driving limit. Interestingly, periodic driving can also conspire to show quantum Kardar-Parisi-Zhang (KPZ)-like superdiffusive dynamical behavior, which seems to have no classical counterpart. We further construct an effective Floquet Hamiltonian, which qualitatively captures this feature occurring in the driven model.
\end{abstract}
\maketitle
\section{Introduction} 
Although the out-of-equilibrium dynamics of closed quantum systems have been studied quite extensively in the past decade~\cite{polkovnikov2011colloquium,outofequilibrium2015,Floquet2019}, the dynamical scaling has not been addressed enough until very recently. In a series of recent theoretical works, a dynamical one-parameter scaling, namely, the Family-Vicsek (FV) scaling is found in a variety of systems, namely, the clean interacting Bose-Hubbard model~\cite{fujimoto2020family}, free fermionic models in presence of random disorder~\cite{fujimoto2021dynamical} and dissipation~\cite{Fujimoto2022dissipation} by studying the surface roughness operator (see Eq.~\ref{surf}). In analogy to the surface growth operator in classical systems, where FV scaling is studied quite extensively~\cite{FVscaling1984,family1985scaling,barabási_stanley_1995}, a similar ``quantum surface-height operator" is introduced in this context~\cite{jin2020stochastic}, which essentially represents the fluctuations of particle-number over a given length scale. The measure of the standard deviation of this operator defines the quantum surface roughness, which plays the key role in giving rise to the one-parameter FV scaling relation. 
The surface growth physics in classical systems primarily demonstrates two dynamical universality classes - the diffusive or Edwards-Wilkinson (EW) class~\cite{EW1982} and the superdiffusive or Kardar-Parisi-Zhang (KPZ) universality class~\cite{KPZ1986} with dynamical exponent $z$~=~2 and 3/2, respectively. Also an anomalous diffusion have been recently reported across the Ising and Kosterlitz-Thouless transitions in classical 2D XXZ model~\cite{ruidas2021}. However, in quantum systems, the most commonly found non-trivial behavior is either diffusive or ballistic. Nevertheless, KPZ-like dynamical behavior of different-time correlations has been found recently in interacting integrable spin chains in presence of the non-abelian symmetries  both theoretically ~\cite{jin2020stochastic,KPZ2019SU2,Ilievski2018,KPZ2022Integrable,keenan2022evidence} and experimentally~\cite{quantumgas2022}. 
As opposed to such different-time correlation functions, quantum surface roughness is a sum of equal time correlations, and can be defined by a local quantity; further, it directly demonstrates FV scaling~\cite{fujimoto2020family}. Moreover, it is an easily accessible quantity using a microscope in cold atomic experiments~\cite{fujimoto2021dynamical}, and provides a direct way of measuring the entanglement entropy in a non-interacting many-body quantum systems~\cite{fujimoto2021dynamical}.
\par Further, out-of-equilibrium dynamics in the presence of quasiperiodic disorder has been a current subject to intense research~\cite{roy2021entanglement,Amna2021,purkayastha2017,varma2017,Iglói_2013}.
Some of the recent works also look into the interplay between periodic driving and quasiperiodic disorder~\cite{sarkar2021mobility,sreemayee22mobility,Sthitadhi18mobility,shimasaki2022anomalous,ray2018}, which give rise to rich phase diagrams with no equilibrium counterparts. Moreover, the recent advances in cold atom experiments enable us to uncover these purely non-equilibrium phases tailored by Floquet engineering in experimental settings~\cite{AAH2015,eckardt2017,messer2018}. Considering the intricacies of out-of-equilibrium dynamics in the quasiperiodic models, one can ask the following questions pertaining to the surface roughness dynamics: Can we find dynamical universality classes other than ballistic and diffusive ones in a simple quasiperiodic non-interacting quantum model?
How good does the FV scaling work for the critical phase having only critical states that lead to critical slowing down of dynamics~\cite{chaikin_lubensky_1995}? Can we capture
the dynamics emerging out of the interplay between quasiperiodic disorder and periodic driving in a one-dimensional closed quantum system within the framework of FV scaling? Is it possible to achieve the emergent KPZ-like dynamics by tuning the parameters appropriately in the driven system?
\par
\begin{figure}
\includegraphics[width=\columnwidth,height=9.0cm]{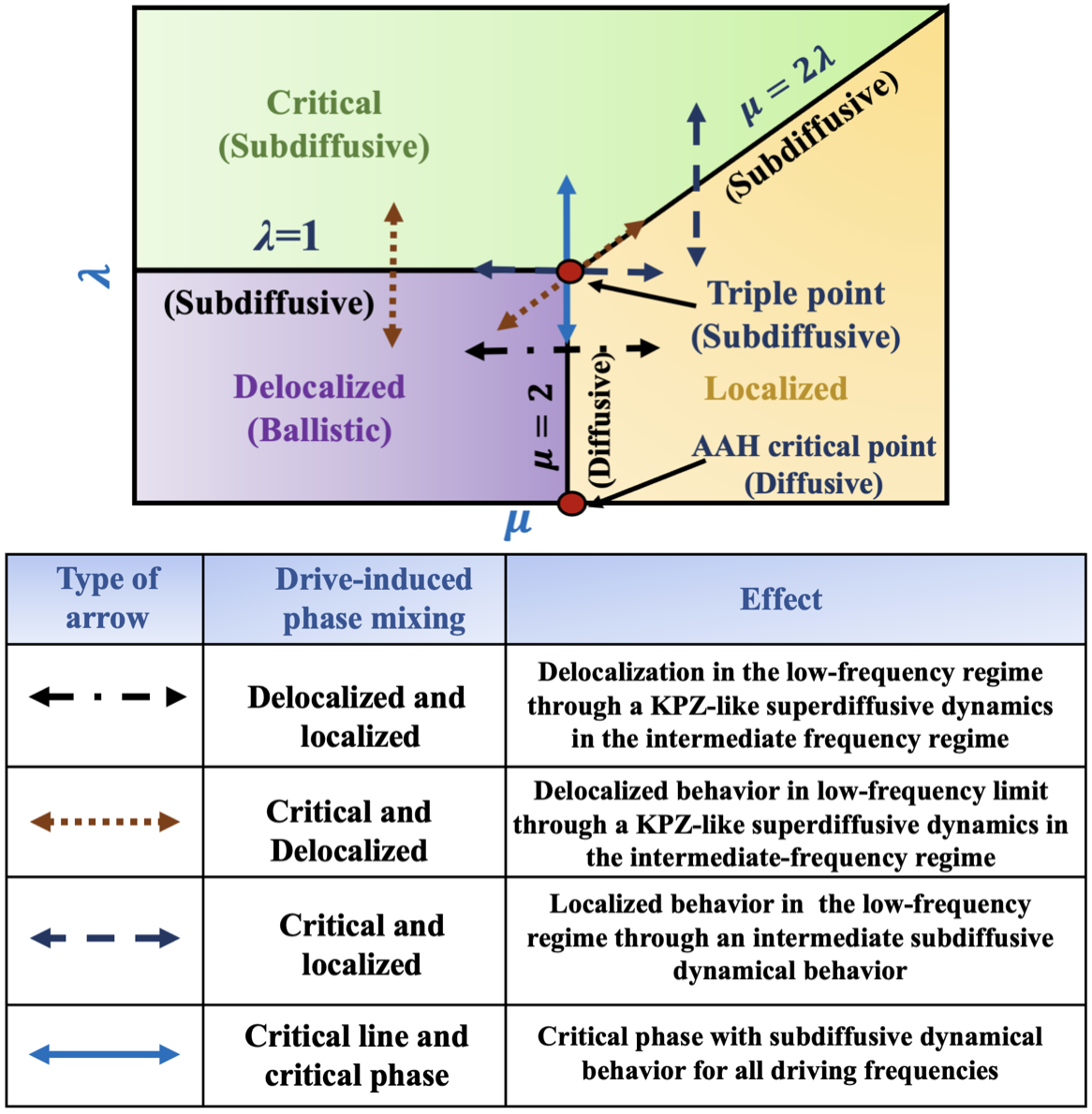}
\caption{{\bf Schematic of the main results:} Top panel: Phase diagram of our static system described in Eq.~\ref{Ham1} along with the direction of mixing when the system is subjected to periodic driving. Bottom panel: A table showing the effect of drive-induced phase mixing in the intermediate to low-frequency regimes.}\label{phase}
\end{figure}

 In this work, we study surface roughness growth in a one-dimensional model of spinless fermions with a quasiperiodic potential and a quasiperiodic hopping in non-equilibrium settings with and without a periodic square pulse protocol. The single-particle phase diagram of the static model, as shown in Fig.~\ref{phase}, comprises three phases: extended, critical, and localized along with three critical lines, which meet at a triple point~\cite{thouless1994critical}. 
We investigate the presence of FV scaling, particularly, on the critical lines and in the critical phase, where the behaviors are expected to be the most intriguing. We find that while the vertical critical line ($\mu=2$) containing the Aubry-Andr\'e critical point ($\lambda=0$, $\mu=2$) demonstrates (almost) diffusive growth dynamics, the other two critical lines ($\lambda=1$ and $\mu=2\lambda$, respectively) and the critical phase exhibit subdiffusive dynamics with anomalous scaling exponents. A scaling analysis for the inverse participation ratio (IPR)~\cite{aoki1986,mirlin2006,JANSSEN19981,evers2008} of the single-particle Floquet eigenstates of the driven system reveals that the (inter-phase) drive-induced mixing between critical and delocalized (localized) phases impels the system toward the delocalization (localization) in the low driving frequency limit. Furthermore, the phase-mixing between delocalized and localized phases shows a re-entrant delocalization transition~\cite{sreemayee22mobility,ray2018,shimasaki2022anomalous} at extremely low driving frequencies. In the stroboscopic dynamics of surface roughness at intermediate driving frequencies, we observe an emergent KPZ-like superdiffusive dynamics, which are primarily governed by the quantum nature~\cite{jin2020stochastic,KPZ2019SU2,Ilievski2018,KPZ2022Integrable,keenan2022evidence} of the model with the states evolving according to the quantum Schr\"odinger equation as well as the quantum nature of the constituent particles (fermions in our case). 
We summarize our results in Fig.~\ref{phase}. Finally, we construct an effective Hamiltonian using the van-Vleck perturbation theory~\cite{Brilloinwigner2016} that describes the KPZ dynamics of the driven model.

The rest of the paper is organized as follows. In Sec. \ref{static model}, we discuss the static model that we consider throughout our work to study the dynamics in the absence and presence of periodic driving. In Sec. \ref{surfaceroughness}, we discuss the dynamical universality classes of the critical lines and critical phase in the framework of Family-Vicsek (FV) scaling while studying the exact surface roughness dynamics in Sec.~\ref{sec3a} and approximate surface roughness dynamics in relation to dynamics of half-chain entanglement entropy in Sec.~\ref{sec3b} starting from a many-body initial state. In addition to that, we also discuss the single-particle wave packet dynamics (charge transport) for the critical lines and critical phase in Sec.~\ref{sec3c}. This is followed by Sec. \ref{periodically driven model} where we examine the effect of inter-phase driving across the critical lines of the static model. Next in Sec. \ref{dynamicsper}, we discuss the dynamical behaviors of the critical phases subjected to periodic driving, which involves Kardar-Parisi-Zhang (KPZ)-like superdiffusive dynamics as detailed in Sec.~\ref{sec5a}. We further explain this emergent behavior in  by employing van Vleck expansion and Floquet perturbation theory in the high-frequency and large driving amplitude regimes in Sec.~\ref{sec5b} and Sec.~\ref{S7}, respectively. This is followed by Sec. \ref{S8}, where we discuss the re-scaled surface height distribution for different dynamical phases appearing in the static model as well as in the driven model. Finally, in  Sec. \ref{conclusion}, we conclude by discussing the main results and point out possible experiments that can confirm our theoretical findings.

\section{The static model}
\label{static model}
We consider a one-dimensional spinless fermionic model with correlated disorders, which is described by the following Hamiltonian
\begin{eqnarray}
H&=&\sum_{j=1}^{L}\left[J+\lambda\cos \left(2\pi\beta \left(j+1/2\right)+\phi\right)\right]\left(c_{j}^{\dagger}c_{j+1}+\rm{H.c.}\right)\nonumber\\
&& 
+\mu\sum_{j=1}^{L}\cos(2\pi\beta j+\phi) n_{j},
\label{Ham1}
\end{eqnarray}
where $J$ and $\lambda$ denote the uniform and quasiperiodic nearest-neighbor hopping, respectively, $\mu$ stands for the quasiperiodic potential, $\phi$ denotes a global phase lying between $0$ and $2\pi$, and $\beta=(\sqrt{5}-1)/2$ is the irrational golden mean. We further set $J$ to be $1$ for the rest of our analysis. The phase diagram of this static model is already well-studied in the literature~\cite{thouless1994critical}, which consists of three phases (extended, critical and localized) and three critical lines separating the phases (see Fig.~\ref{phase}). Interestingly, this model supports no mobility edge. For our future convenience, we use the following terminology while referring to the three critical lines: vertical ($\mu=2$), horizontal ($\lambda=1$), and tilted ($\mu=2\lambda$), respectively, as shown in Fig.~\ref{phase}. In addition to that, these three critical lines meet at a triple point as also indicated in Fig.~\ref{phase}. The other end of the vertical critical line is nothing but the Aubry-Andr\'e (AAH) critical point~\cite{aubry1980}, which is also a self-dual point as already well-known in the literature. Moreover, the vertical critical line also falls under the universality class of critical exponents same as the self-dual point~\cite{SOKOLOFF1985189}. 

\section{The surface roughness and Family-Vicsek scaling}
\label{surfaceroughness}
\subsection{Exact surface roughness dynamics}\label{sec3a}
For dynamical characterization of these phases appearing in the static model, we employ the Family-Vicsek (FV) scaling of surface roughness while specifically focusing on the critical lines and critical phase. 
The main motivation behind the proposal of ``quantum surface roughness" is the correspondence between $\partial_{x}h(x,t)$ and the fluctuation of $\rho(x,t)$ in classical systems, where $h(x,t)$ and $\rho(x,t)$ denote the surface height and the local density, respectively. The correlation function of $\delta\rho(x,t)$ of a one-dimensional nonlinear fluctuating hydrodynamics has earlier been demonstrated to show the same scaling relation as that of $\partial_{x}h(x,t)$ in KPZ equation~\cite{spohn2014nonlinear,mendl2015low,kulkarni2015fluctuating}. The quantum surface-height operator analogous to the classical systems is defined as $\hat{h}_{j}=\sum_{i=1}^{j}(\hat{c}_{i}^{\dagger}\hat{c}_{i}-\nu)$,
where $\nu=N/L$ denotes the filling fraction with $N$ and $L$ being the total number of particles and the length of the system, respectively~\cite{fujimoto2020family,fujimoto2021dynamical,Fujimoto2022dissipation}. Then one can calculate the average surface height, $h_{av}=\frac{1}{L}\sum_{j=1}^{L}Tr(\hat{\rho}(t){\hat{h}_{j}})$,
where $\hat{\rho}(t)$ stands for the density matrix. One can further define the surface roughness $W(L,t)$, which measures the standard deviations of $\hat{h}_{j}$, given by
\bea
W(L,t)=\sqrt{\frac{1}{L}\sum_{j=1}^{L}Tr(\hat{\rho}(t)[\hat{h}_{j}-h_{av}]^{2})}.\label{surf}
\eea
The proposal of FV scaling says that  $W(L,t)$ satisfies the following scaling relation~\cite{FVscaling1984,family1985scaling,fujimoto2020family,fujimoto2021dynamical,Fujimoto2022dissipation}, that is
\bea
W(L,t)=L^{\alpha}~g(t/L^{z}),{\label{FV}}
\eea
where the function $g(x)\sim x^\beta$ and $1$ for $x\ll 1$ and $x\gg 1$, respectively. Hence, Eq.~\eqref{FV} implies that $W(L,t)\propto~t^{\beta}$ for $t\ll t_{sat}$ and $L^\alpha$ for $t\gg t_{sat}$, where the crossover time $t_{sat}=L^z$ with the dynamical exponent $z$ being $\alpha/\beta$. 
As per this scaling relation mentioned above, one should see a data collapse onto a single curve for ordinate $W(L,~t)$ and abscissa $t$ normalized by $L^{\alpha}$ and $L^{z}$, respectively. 
Thus this scaling relation gives rise to two famous dynamical universality classes: the EW class~\cite{antal2008,eisler2013,Hunyadi2004} with $(\alpha=1/2,\beta=1/4,z=2)$ and the KPZ class~\cite{KPZ1986} with $(\alpha=1/2,\beta=1/3,z=3/2)$. Moreover, $(\alpha=1/2,\beta=1/2,z=1)$ corresponds to the ballistic dynamics. In our work, $(\beta<1/4,z>2)$ and $(1/4<\beta<1/2,1<z<2)$ are identified with subdiffusive and superdiffusive dynamical classes, respectively.
\begin{figure}[htbp]
\includegraphics[width=\columnwidth,height=6.5cm]{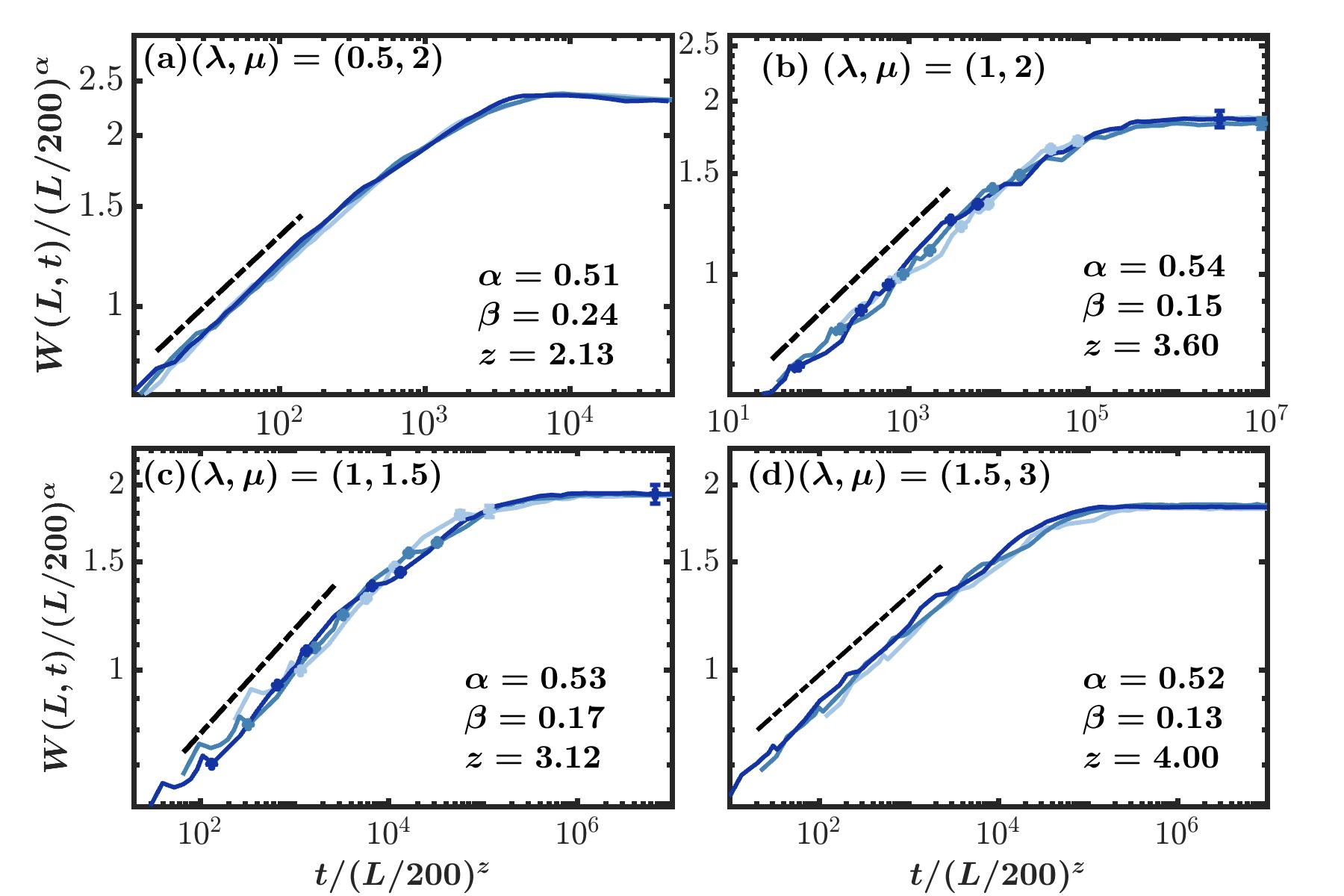}
\caption{{\bf Dynamical scaling of the surface roughness ${\boldmath W (L,t)}$ in the static model:}
(a-d): The variation of $W (L,t)$ with time $t$ normalized by $(L/200)^{\alpha}$ and $(L/200)^{z}$, respectively, for system sizes $L=400,~600$ and $800$. (a): Here we consider the vertical critical line with $\lambda=0.5$, $\mu=~2$. The best fitted scaling exponents for this case are $(\alpha=0.51,\beta=0.24,z=2.13)$, which closely correspond to {\it diffusive/EW} dynamical class. (b-d): The surface roughness dynamics are examined for the triple point ($\lambda=1$, $\mu=2$), horizontal ($\lambda=1$,  $\mu=1.5$), and tilted critical lines ($\lambda=1.5$, $\mu=3$), respectively. The best fitted scaling exponents for these cases are $(\alpha,\beta,z) = (0.54,0.15,3.60)$, $(0.53,0.17,3.12)$, and $(0.52,0.13,4.00)$, respectively, which belong to {\it subdiffusive} dynamical class.} \label{fig01} \end{figure}

\par To study the dynamical behavior of the static model, we consider the many-body initial state $\ket{\Psi_{\text{int}}}=\prod_{j=1}^{L/2} c_{2j}^{\dagger}\ket{0}$ at half-filling, which is  a natural choice~\cite{antal1996Dynamicalscaling,antal2008,eisler2013,Hunyadi2004} due to its easy accessibility in cold atom experiments.
First, we examine the surface roughness dynamics of the vertical critical line ($\lambda=0.5,~\mu=2$) that demonstrate diffusive dynamical behavior with the best fitted scaling exponents $(\alpha,\beta,z)=(0.51,0.24,2.13)$, as  shown in Fig. \ref{fig01} (a).
Further, the AAH critical point ($\lambda = 0$, $\mu = 2$)~\cite{aubry1980} also shows the same exponents as shown in Fig. \ref{exactw} (a), indicating the EW universality class continuing to persist along the vertical critical line. In addition to that, this behavior is consistent with the earlier finding~\cite{thouless1994critical},
which is that the critical exponent $\nu=1$ at self-dual critical point of the AAH model~\cite{aubry1980} continues to be $\nu=1$ along the vertical critical line~\cite{thouless1994critical}, which we confirm through a scaling analysis of entanglement entropy as shown in Figs. \ref{entanglesca} (a-b). We then investigate behavior of the triple point  $(\lambda= 1$, $\mu = 2$), 
where we observe subdiffusive dynamical behavior with the scaling exponents $(\alpha,\beta,z)= (0.54,0.15,3.60)$, as shown in Fig. \ref{fig01} (b).
In Figs.~\ref{fig01} (c-d), the best fitted scaling exponents for the parameter points on the horizontal ($\lambda=1$ and $\mu=1.5$) and tilted critical lines ($\lambda= 1.5,~\mu = 3$) are seen to be $(0.53, 0.17, 3.12)$ and $(0.52, 0.13, 4.00)$, respectively, which indicate the presence of subdiffusive dynamical class. However, the different exponents obtained for the subdiffusive cases are somewhat `anomalous' due to their rare occurrences in the classical systems and clean non-interacting quantum systems. The scaling exponents in the critical phase also lie under the subdiffusive  dynamical class, e.g. at ($\lambda = 1.2$, $\mu = 1.8$) exponents are $(0.53,0.19,2.79)$ as evinced in Fig. \ref{exactw} (b). 
\subsection{Approximated expression of the surface roughness and the relation to the bipartite entanglement entropy}\label{sec3b}
\begin{figure}[h!]
\stackon{\includegraphics[width=0.51\linewidth,height=3.3cm]{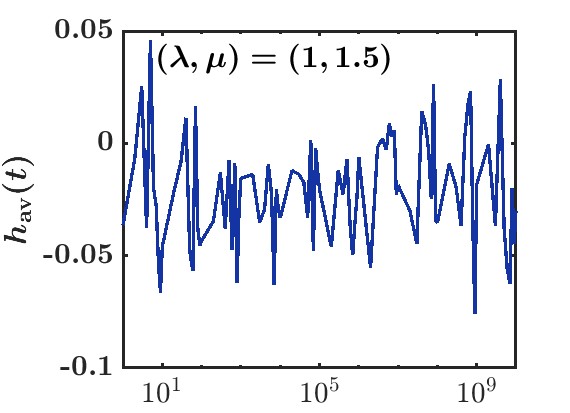}}{(a)}
\stackon{\includegraphics[width=0.48\linewidth,height=3.3cm]{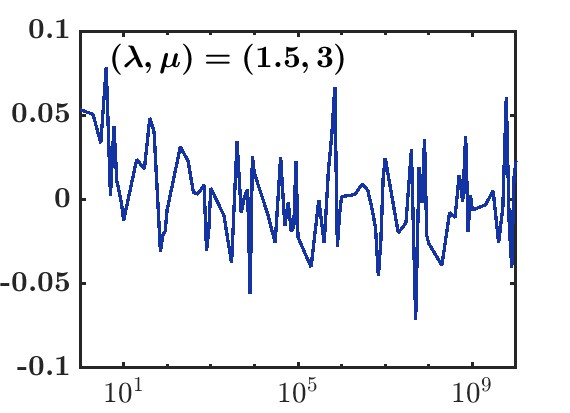}}{(b)}
\stackon{\includegraphics[width=0.51\linewidth,height=3.3cm]{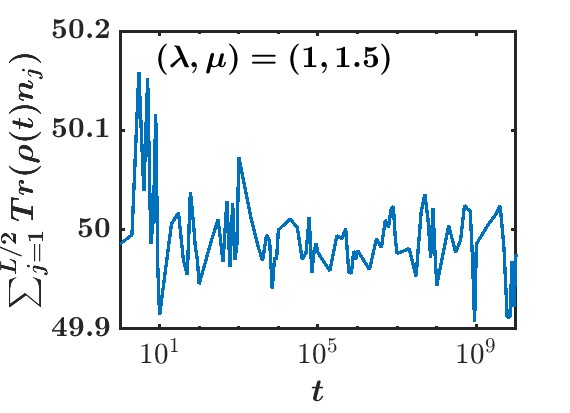}}{(c)}
\stackon{\includegraphics[width=0.48\linewidth,height=3.3cm]{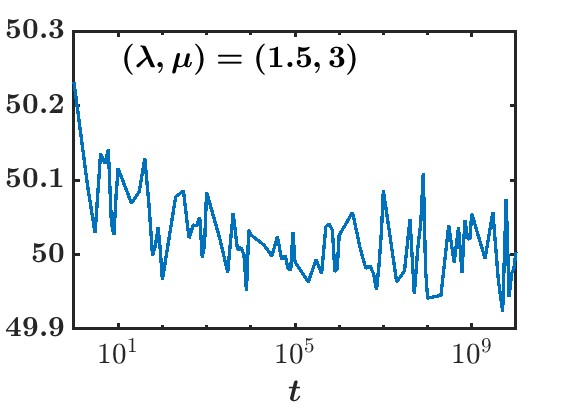}}{(d)}
\caption{{\bf Time evolution of $\mathbf{h_{\rm av}(t)}$ and $\mathbf{\sum_{j=1}^{L/2}}\mathbf{{\rm Tr}(\rho(t)n_{j})}$}: (a-b) The variation of $h_{\rm{av}}(t)$ with time for $L=200$. The parameter values chosen for these plots are as follows: Fig. (a): $\lambda=1,~\mu=1.5$ (a point on the horizontal critical line), Fig. (b): $\lambda=1.5,~\mu=3$ (a point on the tilted critical line). In both cases, $h_{\rm{av}}(t)$ remains close to zero for all time. (c-d) The variation of $\sum_{j=1}^{L/2} {\rm Tr}(\rho(t)n_{j})$ with time for the parameter values same as Figs. (a-b). In both cases, $\sum_{j=1}^{L/2}\rm{Tr}(\rho(t)n_{j})$ remains close to $\frac{\nu L}{2}$, which is $50$ for our case since we have taken the initial state with $\nu=1/2$ with the system size being $L=200$. }
\label{figs111}
\end{figure}

As a further confirmation, we also study the dynamics of the square of the approximate surface roughness $W_a^{2}(L,t)$ and the dynamics of bipartite entanglement entropy $S_{L/2}(t)$. It can be shown that $W_a^{2}(L,t)$ is closely related to the half-chain entanglement entropy $S_{L/2}(L,t)$ \cite{fujimoto2021dynamical} in the case of non-interacting many-body quantum systems provided the following conditions are being satisfied: i) $h_{av}(t)~\approx~0$, ii) $W^{2}(L,t)~\approx~Tr(\hat{\rho}(t)\{h_{L/2}-h_{av}\}^{2})$, and, iii) $\sum_{j=1}^{L/2}Tr(\hat{\rho}(t)n_{j})\approx \frac{\nu L}{2}$, where $h$, $\nu$ stand for the surface-height operator and the filling fraction of the quantum many-body initial state, respectively. We also observe that these quantities obey the previously mentioned conditions for our static model. As example cases,
in Figs. \ref{figs111} (a-d), we  show that the conditions (i) and (iii) are being satisfied for two different sets of parameter values, $\lambda=1,~\mu=1.5$ (a point on horizontal critical line), and $\lambda=1.5,~\mu=3$ (a point on the tilted critical line) for the static model, i.e., $h_{\rm{av}}(t)$ and $\sum_{j=1}^{L/2}{\rm Tr}(\rho(t)n_{j})$ remain close to zero and $\frac{\nu L}{2}$, respectively, for all time.
These assumptions further allow us to define the approximate surface roughness operator in the following manner~\cite{fujimoto2021dynamical}
\bea
W^2(L,t)~\simeq~ W_a^{2}(L,t)~=~Tr\left[\hat{\rho}(t)\left(\sum_{i=1}^{L/2}c_{i}^{\dagger}c_{i}-\frac{L\nu}{2}\right)^{2}\right],\non\\\label{approxw}
\eea
which is equivalent to the particle-number fluctuations in the half of the system. For non-interacting many-body quantum systems, one can also argue that the half-chain entanglement entropy $S_{L/2}(L,t)~$~$\approx~3W_a^{2}(L,t)$~[See supplementary of Ref.~\cite{fujimoto2021dynamical} for a detailed derivation], which implies that the surface roughness can be a good quantifier of entanglement entropy in such systems.

\begin{figure}
\includegraphics[width=\columnwidth,height=6.2cm]{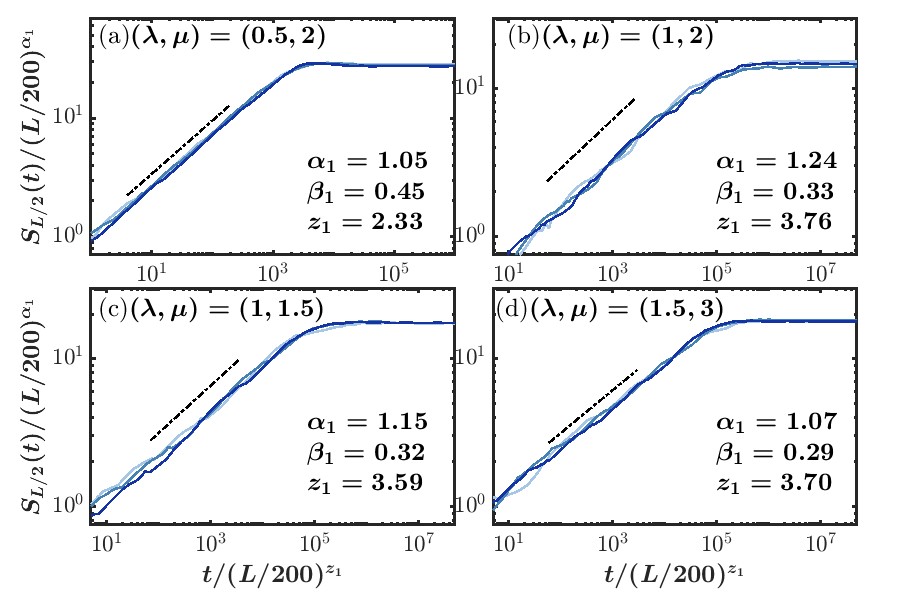}
\caption{{\bf FV scaling of entanglement entropy at criticalities:} 
(a-d) The variation of $S_{L/2}$ with time for $L=400,~600$ and $800$. The parameter values chosen for these plots are as follows: Fig. (a):$\lambda=0.5,~\mu=2$ (vertical critical line), Fig. (b): $\lambda=1,~\mu=2$ (Triple point), Fig. (c): $\lambda=1,~\mu=1.5$ (horizontal critical line), Fig. (d): $\lambda=1.5,~\mu=3$ (tilted critical line). For almost all cases, we see good agreement with FV scaling relations with the scaling exponents $(\alpha_1, \beta_1,z_1)$ being: Fig. Fig. (a): $(1.05,0.45,2.33)$, Fig. (b): $(1.24,0.33,3.76)$, Fig. (c): $(1.15,0.32,3.59)$, Fig. (d): $(1.07,0.29,3.70)$.}\label{figs1}
\end{figure}

\begin{figure}[h!]
\includegraphics[width=\columnwidth,height=6cm]{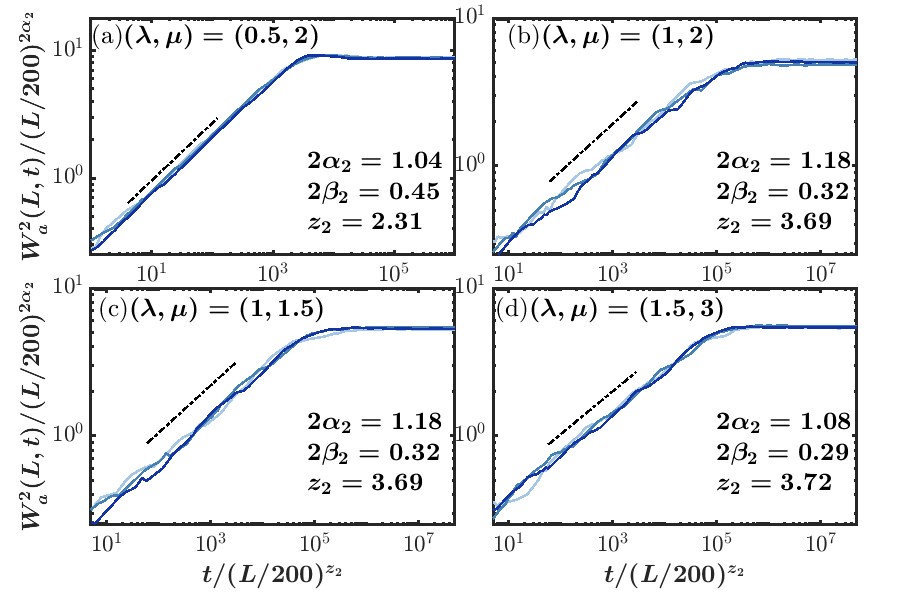}
\caption{ {\bf FV scaling of approximate surface roughness operator $W^{2} (L,t)$ at criticalities:} 
(a-d) Plots of square of the approximate surface roughness $W^{2} (L,t)$  vs $t$ for the system sizes being $L=400,~600$ and $800$. The plot parameters chosen for these figures are the same as Figs. \ref{figs1} (a-d). The best fitted scaling exponents $(2\alpha_2,2\beta_2,z_2)$ estimated from this analysis are as follows: (a) $(1.04, 0.45, 2.31)$, (b) $(1.18,0.32,3.69)$, (c) $(1.18,0.32, 3.69)$, (d) $(1.08,0.29,3.72)$.} \label{figs2} \end{figure}

One can therefore employ FV scaling relations both for $W_a^{2}(L,t)$ and $S(L,t)$ to deduce following scaling relations given below
\bea
S(L,t)~=~L^{\alpha_1}s(t/L^{z_1})&~\propto~&~t^{\beta_1}~~~t<<t_{sat},\non\\
&~\propto~&~L^{\alpha_1}~~~t>>t_{sat}.\non\\
W_a^{2}(L,t)=L^{2\alpha_2}w(t/L^{z_2})&~\propto~&~t^{2\beta_2}~~~t<<t_{sat},\non\\
&~\propto~&~L^{2\alpha_2}~~~t>>t_{sat}.\non\\\label{FV2}
\eea
Similar to the one-parameter scaling relation for the exact surface roughness operator $W(L,t)$ shown in Eq. (\ref{FV}), one can further argue that the crossover time for these cases satisfies $t_{sat}=L^{z_1 (z_2)}$, where $z_{1(2)}=\alpha_{1(2)}/\beta_{1(2)}$. Hence, one should also expect a scaling collapse of $S(L,t)$ ($W_a^{2}(L,t)$) vs $t$, followed by a proper normalization of ordinate and abscissa by $L^{\alpha_1}$ ($L^{2\alpha_2}$) and $L^{z_1}$ ($L^{z_2}$), respectively. As shown in Fig. (\ref{figs1}), we numerically investigate the dynamics of the bipartite entanglement entropy $(S_{L/2})$  for the same density-wave (DW) initial state to confirm the exponents obtained from the scaling analysis of the exact surface roughness dynamics. In Fig. \ref{figs1} (a), we consider the vertical critical line ($\lambda=0.5$ and $\mu=2$), where we observe diffusive dynamical behaviors with $(\alpha_1,\beta_1,z_1)=(1.05,0.45,2.33)$, respectively. In addition to that, we see a similar behavior for the AAH citical point ($\lambda=0,~\mu=2$). In Figs. \ref{figs1} (b-d), we again consider four parameter points on the critical phase boundaries, namely, the triple point ($\lambda=1,~\mu=2$), horizontal critical line ($\lambda=1,~\mu=1.5$) and tilted critical line ($\lambda=1.5,~\mu=3$). In all three cases, the system demonstrates subdiffusive dynamical behaviors with the best fitted scaling exponents being $(\alpha_1,\beta_1,z_1)=(1.24,0.33,3.76)$, $(1.15,0.32,3.59)$, and $(1.07,0.29,3.70)$, respectively. In addition to that, we observe the critical phase demonstrating similar subdiffusive dynamical behavior (not shown here). These scaling exponents more or less agree with the scaling exponents obtained from the exact $W(L,t)$ dynamics.
In Figs. \ref{figs2} (a-d), we further numerically examine the dynamics of $W_a^{2}(L,t)$ as defined in Eq. \eqref{approxw} for the same many-body initial state, which essentially carries the information about the particle-number fluctuation in the half of the system. As shown in Figs. \ref{figs2}, the best fitted exponents obtained from this analysis for the same sets of parameter values  are also found to be close to those obtained from the previous two dynamical measurements. 
\subsection{single-particle wave packet dynamics in the static model}\label{sec3c}
\begin{figure}[h!]
\includegraphics[width=\columnwidth,height=6.2cm]{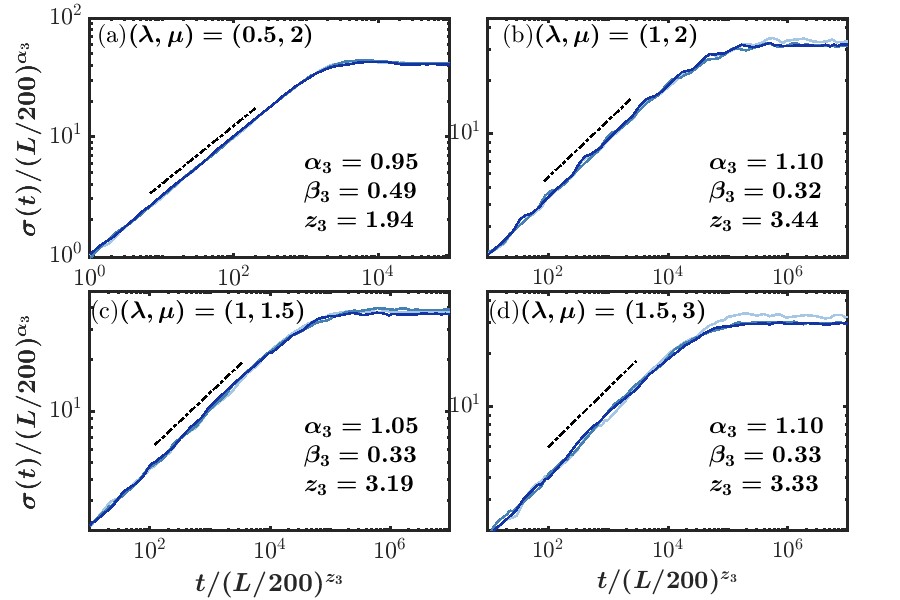}
\caption{ {\bf FV scaling from the single particle wave packet dynamics:}
(a-d) The dynamics of the single-particle root-mean-squared displacement $(\sigma(t)=\sqrt{\langle X^{2}(t)\ra-\la X(t)\ra^2})$ vs $t$ for system sizes $L=400,~600$ and $800$. The parameter values fo these plots are the same as Fig. \ref{figs1} (a-d). The single-particle dynamics also respect the FV scaling relations as seen in Figs. (a-d) with the best fitted scaling exponents $(\alpha_3,\beta_3,z_3)$ being : Fig(a): $(0.95,0.49,1.94)$, Fig(b): $(1.10,0.32,3.44)$, Fig(c): $(1.05,0.33,3.19)$, Fig(d): $(1.10,0.33,3.33)$.}\label{figs3}
\end{figure}

In addition to the many-body DW dynamics, we also examine the dynamics of the single-particle root-mean-squared displacement (RMSD), which is given by
\bea
\sigma(t)&=&\sqrt{\langle X^{2}(t)\rangle-\langle X(t)\rangle^2},\non\\
\langle X^{2}(t)\rangle &=&\sum_{j=1}^{L}(j-j_{0})^{2}~|\psi_{j}(t)|^{2},\non\\
\langle X(t)\rangle &=& \sum_{j=1}^{L}(j-j_{0})~|\psi_{j}(t)|^{2},\label{msd}
\eea
where the initial state is a single-particle wave packet localized at site $j_{0}$, and $\psi_{j}(t)$ is the $j$-th component of the time evolved initial state with the static Hamiltonian shown in Eq. \ref{Ham1}.  In Figs. \ref{figs3} (a-d), we examine the wave packet dynamics at the critical phase boundaries, namely, vertical critical point ($\lambda=0.5,~\mu=2$), triple point ($\lambda=1,~\mu=2$), horizontal critical line ($\lambda=1,~\mu=1.5$), and tilted critical line ($\lambda=1.5,~\mu=3$). The scaling exponents ($\alpha_3,~\beta_3,~z_3$) found from FV scaling for these cases are found to be $(0.95,0.49,1.94)$, $(1.10,0.32,3.44)$, $(1.05,0.33,3.19)$ and $(1.10,0.33,3.33)$, respectively, which also respect the FV scaling relation. Note that the first dynamical exponents again show diffusive dynamical behaviors as seen in the previous cases. On the other hand, the last three dynamical exponents lie in the subdiffusive dynamical regimes, which agree with the behaviors obtained from different dynamical measurements performed with the many-body DW initial state. We have checked that the many-particle RMSD calculated for DW intial state also produces same exponents (not shown here).

From our numerical analysis we find that $\beta_1=2\beta_2\approx\beta_3$ and $\alpha_1=2\alpha_2\approx\alpha_3$. This shows the equivalence between the growths of entanglement ($S_{L/2}(t)$) and number fluctuations ($W_a^{2}(L,t)$) and also establishes a relationship between surface roughness growth and transport exponents. A similar relationship has been reported in a recent study for an interacting quasi-periodic model~\cite{bhakuni2023dynamic}. Although for some instances, it has been observed that the transport exponents show significant disagreement with the surface roughness exponents, such as in the random-dimer model (RDM)~\cite{fujimoto2021dynamical}. Hence, this relationship as we obtain for our model, may not be a universal one and can be model-dependent. But this needs more serious investigations in the future. 

\section{Periodically driven model and its phase diagram}
\label{periodically driven model}
For the driven case, we specifically focus on the inter-phase driving across the critical lines to systemically explore the phase diagram. To do so, we drive the static Hamiltonian with a square pulse protocol in the following manner
\bea
H(t)&=& H(\lambda+\delta\lambda~f(t),\mu+\delta\mu~f(t)),
\label{Ham_drive}
\eea
where $f(t)=1~(-1)$ for $0\leq t \leq T/2$ $(T/2\leq t\leq T)$; $\delta\lambda$ and $\delta\mu$ denote the amplitudes of periodically driven quasiperiodic hopping and onsite potential, respectively. $T=2\pi/\om$ denotes the time period for the stroboscopic time evolution, and $\om$ is the driving frequency. For such a square pulse protocol, the Floquet evolution operator is given by, $U_{F}(T,0)$~=~$e^{-iH_{2}T/2}e^{-iH_{1}T/2}$~=~$\sum_{m}e^{-i\theta_{m}}\ket{m}\bra{m}$~\cite{BLANES2009151,Sen_2021,burkov2015},
where $H_1$ ($H_2$) denotes the Hamiltonian for $0\leq t \leq T/2$ ($T/2\leq t\leq T$). $e^{-i\theta_{m}}$ is the $m$-th Floquet eigenvalue, and $\ket{m}=\sum_{j}\psi_{m}(j)\ket{j}$ defines the $m$-th single particle Floquet eigenstate with $j$ denoting the coordinate of the lattice sites.
To study the localization properties of a normalized Floquet eigenstate, we define the inverse participation ratio, given by, $I_{m}^{(2)}$ = $\sum_{j}|\psi_{m}(j)|^{4}$~\cite{aoki1986,mirlin2006,JANSSEN19981,evers2008}.
The quantity $I_{m}^{(2)}$ scales with system size $L$ as $L^{-\eta_{2}}$, where $\eta_{2}$ is the (multi)fractal dimension. $\eta_{2}=0$ (1) for localized (extended) states whereas $\eta_{2}$ should lie between $0$ and $1$ for critical states~\cite{evers,caste}.

\begin{figure}[h!]
\includegraphics[width=\columnwidth,height=7cm]{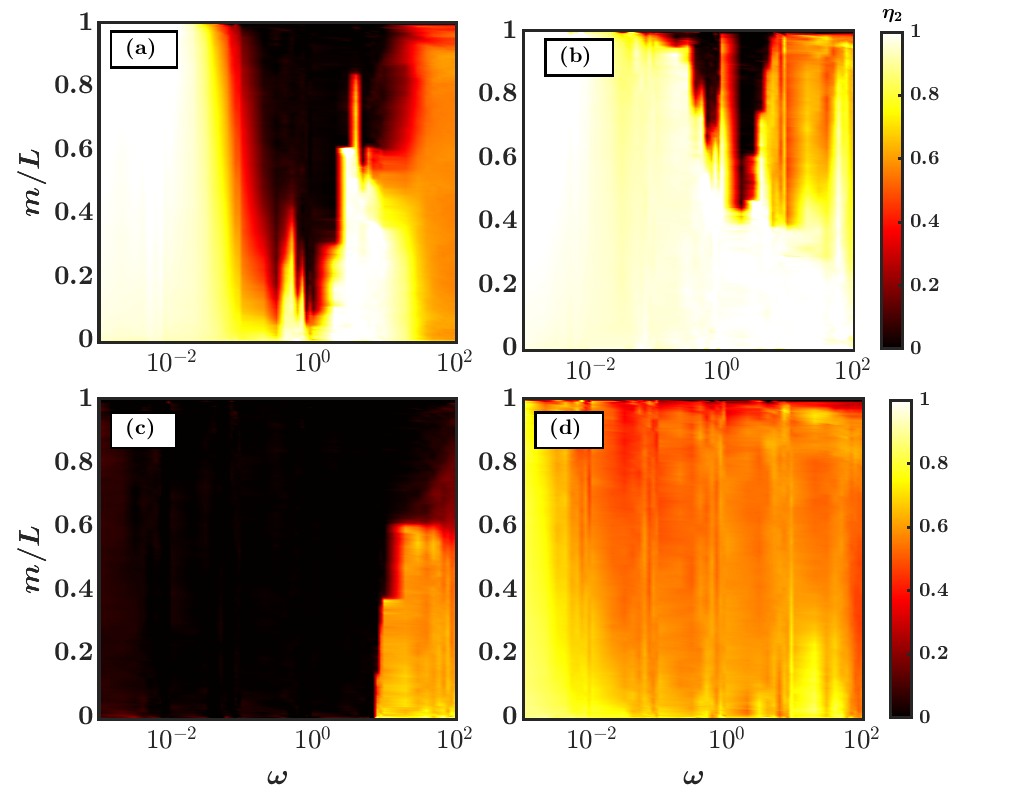}
\caption{{\bf Fractal dimension $\eta_2$ of the single-particle  Floquet eigenstates:} (a-d) The variation of $\eta_{2}$ 
as a function of the fractional single-particle Floquet eigenstate index, $m/L$ (where $m$ denotes the eigenstate index after arranging total $L$ Floquet eigenstates in the ascending order of their IPR values), and frequency $\omega$ for $L=500,~1000$, $2000$, $3000$ and $4000$. (a): $\lambda=0.5$, $\mu=2$, $\delta\lambda=0$, and $\delta\mu=1.5$ (Delocalized and localized phase mixing across the vertical critical line). (b): $\lambda=1$, $\mu=1.5$, $\delta\lambda=0.8$, and $\delta\mu=0$ (Critical and delocalized phase mixing across the horizontal critical line). (c): $\lambda$=1.5, $\mu=3$, $\delta\lambda$~=~0, and $\delta\mu=2.5$ (Critical and localized phase mixing across the tilted critical line). (d): $\lambda=1.0$, $\mu=2.0$, $\delta\lambda=0.5$, and $\delta\mu=0$ (Critical line and critical phase mixing across the triple point).} \label{fig06} \end{figure}

As shown in Fig. \ref{fig06} (a-d), we examine the phase diagram of the periodically model, driven across the critical lines, by extracting $\eta_{2}$ of the single-particle spectrum-resolved Floquet eigenstates (sorted in the increasing order of IPR-values), as a function of the fractional eigenstate index, $m/L$, and driving frequency, $\omega$.
In Fig. \ref{fig06} (a), we first consider the vertical critical line ($\lambda = 0.5$, $\mu = 2$) of the static model, and allow inter-phase driving between delocalized and localized states with $\delta\lambda=0$ and $\delta\mu=1.5$. We see the Floquet eigenstates exhibiting critical behavior in the high-frequency regime just like the static case, coexisting phases in the intermediate-frequency regime, and finally  delocalized behavior~\cite{ray2018,sreemayee22mobility,shimasaki2022anomalous} at extremely small $\om$-values. In Fig. \ref{fig06} (b), we show results for the horizontal critical line ($\lambda=1,~\mu=1.5$) for drive-induced mixing between delocalized and critical regimes with $\delta\lambda$~=~0.8 and $\delta\mu=0$. In this case, we observe that the Floquet eigenstates first exhibit critical behavior in the high-frequency limit (same as the static model), which, after showing intriguing features at intermediate frequencies, gradually moves toward complete delocalization with decreasing $\omega$-values in the slow driving limit. In Fig. \ref{fig06} (c), the tilted critical line ($\lambda=1.5$ and $\mu=3$) is subjected to periodic driving along the horizontal direction with $(\delta\lambda,~\delta\mu)$~=~$(0,2.5)$, which only allows mixing between critical and localized phases. In this case, the critical regime in the high-frequency regime (like the static case) slowly flows towards the complete localization in the comparatively slow driving limit. 
Finally, we drive the triple point $(\lambda,\mu)=(1,2)$ along the vertical direction with $(\delta\lambda,\delta\mu)=(0.5,0)$, which gives rise to only critical Floquet eigenstates, as shown in Fig.~\ref{fig06} (d), for all driving frequencies due to the drive-induced phase-mixing between critical line (vertical) and critical phase. 
We then perform similar analysis for several other choices of parameters (not shown here) and summarize our general findings about the drive-induced phases in the table shown in Fig.~\ref{phase}. Our analyses establish that the critical phase is quite fragile when it competes for its existence with delocalized or localized phase via periodic driving. Moreover, inter-phase driving between delocalized and localized phases in general flows towards complete delocalization in the low-frequency regime~\cite{ray2018,sreemayee22mobility,shimasaki2022anomalous}, followed by an intricate intermediate-frequency regime, which we will discuss later.
We thus reach a hierarchy of phases in terms of the stability in the slow driving limit, i.e, $\text{delocalized}>\text{localized}>\text{critical}$.

\section{Dynamical scaling in the periodically driven model}
\label{dynamicsper}
\subsection{Surface roughness dynamics in the periodically driven model}\label{sec5a}
\begin{figure}[htbp]
\includegraphics[width=\columnwidth,height=6.5cm]{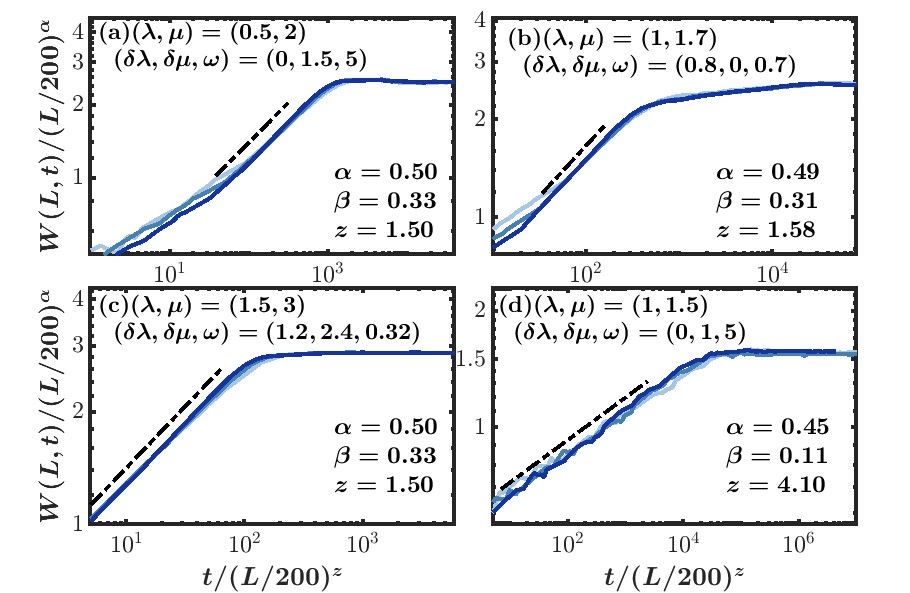}
\caption{{\bf Superdiffusive and subdiffusive dynamics of surface roughness caused by drive-induced mixing between different phases:} (a-d): The stroboscopic dynamics of $W(L,t)$ with time $t$ normalized by $(L/200)^{\alpha}$ and $(L/200)^{z}$, respectively, for $L=400,~600$ and $800$. (a): The vertical critical line $(\lambda=0.5,~\mu=2)$ is considered to allow drive-induced mixing between delocalized and localized states with $(\delta\lambda,~\delta\mu,\om)=(0,1.5,5)$. (b-c): The horizontal $(\lambda=1,\mu=1.7)$ and tilted critical lines $(\lambda,\mu)=(1.5,3)$ are subjected to periodic driving with $(\delta\lambda,\delta\mu,\om)=(0.8,0,0.7)$ and $(1.2,2.4,0.32)$, respectively, to intermix critical and delocalized states. (d): The horizontal critical $(\lambda=1,\mu=1.5)$ is driven with $(\delta\lambda,\delta\mu,\om)=(0,1,5)$, which causes the critical and localized phase to intermix with one another. In Figs. (a-c), we observe KPZ-like {\it superdiffusive} surface roughness dynamics with the best fitted scaling exponents, $(\alpha,\beta,z)$~=~$(0.50,0.33,1.50)$, $(0.49,0.31,1.58)$, and $(0.50,0.33,1.50)$, respectively. In Fig. (d), we see {\it subdiffusive} dynamics with scaling exponents $(\alpha,\beta,z)$ = $(0.45,0.11,4.10)$.}\label{fig05}
\end{figure}
After exploring the IPR phase diagram, we then examine the stroboscopic dynamics of $W(L,t)$~\cite{fujimoto2021dynamical} in the driven model for the same initial many-body state. At first, we analyze the inter-phase mixing between delocalized and critical/localized states across the critical lines, which gives rise to coexisting phases in the Floquet spectrum in the intermediate-frequency regime, as shown in Figs. \ref{fig06} (a-b). This regime is the most intriguing from the dynamical perspective since 
drive-induced subdiffusive and ballistic surface roughness growth in the high-frequency and low-frequency regime, respectively, are already present in the static model due to presence of critical and delocalicalized states. Hence, the most suitable candidates for drive-induced superdiffusive behavior (not present in static model) are those where the multiple phases can coexist in the spectrum with a significant fraction of states being delocalized. This may lead to suppression of the ballistic growth of delocalized states due to the slow dynamics of localized/ critical states, and the net result can be superdiffusive in nature~\cite{nair2023emergent}.
Noting this fact, in Figs. \ref{fig05} (a-c), the vertical ($\lambda=0.5$, $\mu=2$), horizontal ($\lambda=1$, $\mu=1.7$), and tilted $(\lambda,\mu)=(1.5,3)$ critical lines are subjected to periodic driving with $(\delta\lambda,~\delta\mu,~\om)=(0,1.5,5)$, $(0.8,0,0.7)$ and $(1.2,2.4,0.32)$, respectively, to allow phase mixing between either delocalized and localized phases (vertical) or delocalized and critical phases (other two critical lines); further, Floquet spectrum for all cases embrace the property mentioned earlier. The best fitted $(\alpha,\beta,z)$ for these cases are found to be $(0.50,0.33,1.50)$, $(0.49,0.31,1.58)$, and $(0.50,0.33,1.50)$, respectively, which implies an emergent KPZ-like superdiffusive dynamics. 
Since the delocalized and critical phases of the static model exhibit ballistic and subdiffusive behaviors, phase mixing can, in principle, give rise to superdiffusive dynamics, even with KPZ scaling exponents if driven appropriately, which we observe from Fig. \ref{fig05} (b-c). Similar argument also holds for mixing between delocalized and localized states, as shown in Fig. \ref{fig05} (a). 
\par We now discuss the surface roughness growth for phase mixing between critical and localized phases.
As evinced in Fig. \ref{fig05} (d), the horizontal critical line ($\lambda=1,~\mu=1.5$) is subjected to periodic driving with $(\delta\lambda,\delta\mu,\om)=(0,1,5)$, which allows the  phase mixing between critical and localized phases in the intermediate-frequency regime. In this case, we observe subdiffusive behavior~\cite{lezma2019} with $(\alpha,\beta,z)=(0.45,0.11,4.10)$, where the subdiffusive growth exponent of the static model ($\beta\sim0.15$) gets further suppressed due to the presence of the coexisting localized states in the spectrum. Further, the fate of such phase mixing in the extremely low-frequency regime is mostly complete localization with no well-defined FV scaling exponents, as shown in Figs. \ref{figs4}.
whereas in the extremely fast driving limit the scaling exponents agree with the exponents obtained for the static case as shown in Figs \ref{figs6}.

\subsection{KPZ-like scaling behavior on the vertical critical line subjected to periodic driving and the effective van Vleck Hamiltonian}\label{sec5b}
\label{S6}
To get an analytical insight into the KPZ-like scaling behavior, we consider a point on the vertical critical line of the static model with $\lambda=0.5,~\mu=2$, and periodically drive it along the horizontal direction of the static phase diagram with $\delta\lambda=0$, $\delta\mu=1.5$ and $\om=5$ (Fig.~4(a) in the main text). This is comparatively fast driving regime, however not extremely fast so that the signatures of periodic driving can still be present in the stroboscopic Floquet dynamics. These parameter values allow us to use the van Vleck perturbative expansion~\cite{Brilloinwigner2016} to find an effective Floquet Hamiltonian. Before proceeding further, one should note that the stroboscopic Floquet evolution operator for the square pulse protocol, shown in the main text, possesses the following symmetry property, which will be important later in the discussion,
\bea
U_{F}^{-1}(J,\lambda,\mu,\delta\mu,\delta\lambda)=U_{F}(-J,-\lambda,-\mu,\delta\mu,\delta\lambda). \label{sym}
\eea
This further implies the following symmetry property for the effective Floquet Hamiltonian
\bea
H_{F}(J,\lambda,\mu,\delta\mu,\delta\lambda)=-H_{F}(-J,-\lambda,-\mu,\delta\mu,\delta\lambda). \label{sym}
\eea
We can now find the Floquet Hamiltonian $H_F$ in the relatively high-frequency regime 
employing the van Vleck perturbation theory. Note that this is a 
perturbative expansion in powers of $1/\om$, and it has the advantage over 
the Floquet-Magnus expansion that it does not depend on the phase of the
driving protocol, i.e., it is invariant under $f(t) \to f(t+t_0)$. The high-frequency expansion obtained from the van Vleck perturbation theory to the third order~\cite{Brilloinwigner2016} is as follows:
\begin{widetext}
\bea
H_{F}&=&\sum_{n=0}^{\infty}H_{F}^{(n)},\non\\
H_{F}^{(0)}&=&H_{0},\non\\
H_{F}^{(1)}&=&\sum_{m\neq 0}\frac{[H_{-m},H_{m}]}{2m\om},\non\\
H_{F}^{(2)}&=&\sum_{m\neq0}\frac{[[H_{-m},H_0],H_m]}{2m^2\om^2}\,+\,\sum_{m\neq0}\sum_{n\neq0,m}\frac{[[H_{-m},H_{m-n}],H_n]}{3mn\om^2},\non\\
H_F^{(3)}&=&\sum_{m\neq0}\frac{[[[H_{-m},H_0],H_0],H_m]}{2m^3\om^3}+\sum_
{m\neq0}\sum_{n\neq0,m}\frac{[[[H_{-m},H_0],H_{m-n}],H_n]}{3m^2n\om^3}\non\\
&&+\sum_{m\neq0}\sum_{n\neq0,m}\frac{[[[H_{-m},H_{m-n}],H_{0}],H_{n}]}{4mn^2\om^3}+\sum_{m,m\neq0}\frac{[[[H_{-m},H_{m}],H_{-n}],H_{n}]}{12mn^2\om^3}\non]\\
&&+\sum_{m\neq0}\sum_{n\neq0,m}\frac{[[H_{-m},H_{0}],[H_{m-n},H_{n}]]}{12m^2n\om^3}+\sum_{m,n\neq0}\sum_{l\neq0,m,n}\frac{[[[H_{-m},H_{m-l}],H_{l-n}],H_n]}{6lmn\om^3}\non\\
&&+\sum_{m,n\neq0}\sum_{l\neq 0,m-n}\frac{[[[H_{-m},H_{m-n-l}],H_{l}],H_n]}{24lmn\om^3}+\sum_{m,n\neq0}\sum_{l\neq0,m,n}\frac{[[H_{-m},H_{m-l}],[H_{l-n},H_{n}]]}{24lmn\om^3},
\eea
\end{widetext}
where $H_m$ stands for the $m$-th Fourier component of $H(t)$. The zeroth-order Floquet effective Hamiltonian given by van Vleck perturbation theory for our case is
\begin{widetext}
\bea H_{F}^{(0)} &=& \frac{1}{T} ~\int_0^T dt ~H(t)
= \sum_{j} [(1+\lambda \cos(2\pi\beta(j+1/2)+\phi)~c_{j}^{\dagger}c_{j+1}+ {\rm H.c.}] + \mu \sum_j \cos(2\pi\beta j+\phi) ~c_{j}^{\dagger}c_{j}. \label{H0}
 \eea
\end{widetext}
Now, the $m$-th Fourier component of $H(t)$ turns out to be
\bea H_{m}&=&\frac{2i \delta\mu}{m\pi} \sum_j \cos(2\pi\beta j+\phi)c_{j}^{\dagger}c_{j},
~~ {\rm for} ~~ m ~~\text{odd},\non\\
&=& 0,~~~~ {\rm for} ~~~~~~~~~ m ~~\text{even}, \label{FT}\eea
and $H_0 = H_F^{(0)}$.
We then find that the first-order Floquet Hamiltonian, which is given by $H_F^{(1)} = \sum_{m 
\neq 0}\left[H_{-m},H_{m}\right]/(2m\om)$, and is zero 
for all $m$. The second-order Floquet Hamiltonian 
consists of two terms
\bea H_{F}^{(2)}&=&\sum_{m\neq0}\frac{[[H_{-m},H_{0}],H_{m}]}{2m^{2}\om^{2}}\non\\&&
+\sum_{m\neq0}\sum_{n\neq0}\frac{[[H_{-m},H_{m-n}],H_{n}]}{3mn\om^{2}}. 
\label{hf2} \eea
From our calculation, $H_{F}^{(2)}$ is found to be

\bea
H_{F}^{(2)}&~=~&-\frac{\delta\mu^{2}\sin^{2} (\pi\beta)}{4\om^2}\sum_{j}\sin^{2}(\pi\beta(2j+1)+\phi)\times\non\\&&\left(1+\lambda\cos(2\pi\beta(j+1/2)+\phi)\right)c_{j}^{\dagger}c_{j+1}+\rm{H.c}.\non\\ \label{HF2}
\eea
Therefore, the effective Hamiltonian to second order in $(1/\om)$ will be $H_{F}=H_{F}^{(0)}+H_{F}^{(2)}$. We further note that the third order effective Hamiltonian appears to be zero for our model for this particular choice of the driving protocol, which can be shown by the following argument. The general expression of the third order van Vleck effective Hamiltonian contains multiple terms; however, only two terms which are important for our calculation are as follows (the rest of the terms immediately go to zero due to the fermionic commutation relations),
\bea
H_{F}^{(3)}&=&\sum_{m\neq0}\frac{[[[H_{-m},H_{0}],H_{0}],H_{m}]}{2m^3 \om^3}\non\\&&+\sum_{m\neq0,~n\neq0,~m}\frac{[[[H_{-m},H_{0}],H_{m-n}],H_{n}]}{3m^2 n \om^3}.\non\\
\label{HF3}
\eea
Moreover, the first term shown in Eq. \eqref{HF3} can be shown to be zero due to the symmetry property discussed in Eq. \eqref{sym}~\cite{sreemayee22mobility}. Furthermore, the second term also reduces to zero using the fact that the Fourier components, $H_{m}$, $H_{n}$ and $H_{m-n}$ simultaneously can not be  non-zero due to the constraints imposed by the Fourier expansion, as shown in Eq. \eqref{FT}~\cite{sreemayee22mobility}. Therefore, the effective Hamiltonian to second order in $1/\om$ is a good approximation to the exact Floquet effective Hamiltonian.
With this effective Hamiltonian, we can now study the surface roughness dynamics stroboscopically in time.
\begin{figure}[h!]
\includegraphics[width=\columnwidth,height=5.3cm]{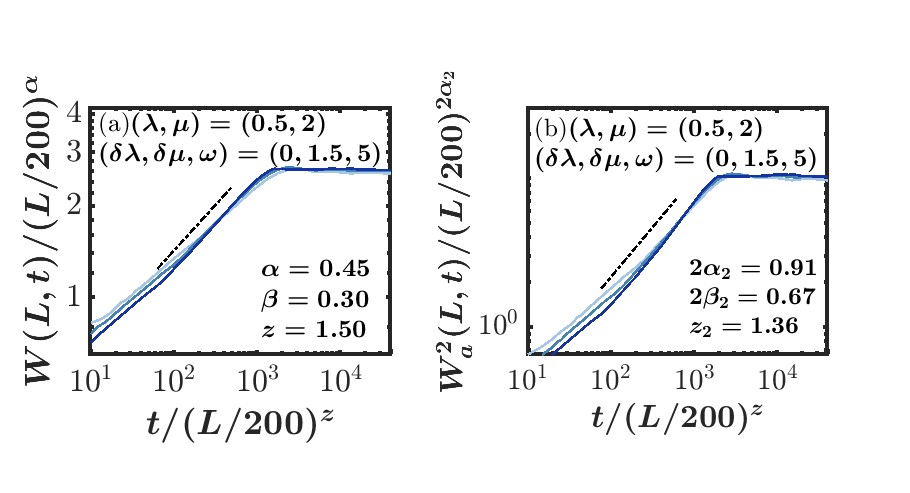}
\caption{{\bf KPZ-like growth from the effective van Vleck Hamiltonian:} (a) The variation of exact $W(L,t)$ in time for the $\lambda=0.5,~\mu=2,~\delta\mu=1.5,~\om=5$, (b) The variation of approximate $W_a^2(L,t)$ in time for the same set of parameter values. The estimated FV scaling exponents are found to be as follows: Fig. (a): $(\alpha,~\beta,~z)=(0.45,~0.30,~1.50)$, Fig. (b): $(2\alpha_2,~2\beta_2,~z_2)=(0.91,~0.67,~1.36)$.} \label{fig10} 
\end{figure}
In Fig. \ref{fig10} (a-b), we analyze the variation of exact surface roughness and the square of the approximate surface roughness operators, $W(L,t)$ and $W_a^{2}(L,t)$ vs time, for a point on the vertical critical line of the static model subjected to periodic driving with parameter values, $\lambda=0.5,~\mu=2,~\delta\mu=1.5$, and $\om=5$ with the effective Floquet Hamiltonian, shown in Eq. \eqref{H0} and Eq. \eqref{HF2}. Remarkably, the best fitted scaling exponents obtained from these analysis for these parameter values are found to be $(\alpha,~\beta,~z)=(0.45,~0.30,~1.50)$, and $(2\alpha_2,~2\beta_2,~z_2)=(0.91,~0.67,~1.36)$, respectively, which show the superdiffusive surface roughness dynamics with close KPZ-like scaling behavior. Moreover, these scaling exponents are quite close to the exponents obtained from exact numerical Floquet analysis, revealing that this effective Floquet Hamiltonian obtained from van Vleck expansion is able to capture the feature of the exact numerical Floquet analysis. 
\subsection{Effective Floquet Hamiltonian in large driving amplitude limit}
\label{S7}
In this section, we will discuss both the IPR-based phase diagram and dynamical behaviors of the periodically driven model in the large driving amplitude limit.
To do so, we will employ the Floquet perturbation theory (FPT) \cite{soori2010, Sen_2021} to find an effective Floquet Hamiltonian for $\delta\mu \gg J,~\mu,~\lambda$. We further choose $\delta\lambda=0$ since this choice of parameter values are easy to handle for our analytical treatment. For this particular case, the unperturbed Hamiltonian, $H_{0}$ is as follows
\bea
H_{0}=\delta\mu f(t) \sum_{j}\cos(2\pi\beta j+\phi)n_{j},\label{unp}
\eea
which is diagonalizable in real space having the instantaneous energy eigenvalues, $E_{j}=\delta\mu f(t)\cos(2\pi\beta j+\phi)$ in the single-particle limit, where $j$ stands for real space lattice indices ranging between 1 and $L$. These instantaneous eigenvalues satisfy the following condition: 
\bea e^{i \int_{0}^{T}dt \left[E_{j}(t) - E_{j'} (t)\right]} =1, \eea
where $T = 2 \pi/\om$ is the driving period. This condition implies that one needs to carry out degenerate FPT~\cite{Sen_2021,soori2010} to find the effective Floquet Hamiltonian. The eigenfunction corresponding to a  particular $E_{j}$ in the single-particle limit is given by 
\bea
\ket{j}=\hat{c}_{j}^{\dagger}\ket{0}.
\eea
To chart out the main steps of FPT for our periodically driven model, we begin with the Schr\"odinger equation 
\beq i\frac{d\ket{\psi}}{dt} ~=~ \left( H_{0}+V \right)\ket{\psi},
\label{schr1} \eeq 
where $V$ is given by
\bea
V&=&\sum_{j}\left(1+\lambda\cos(2\pi\beta (j+1/2)+\phi)\right)~c_{j}^{\dagger}c_{j+1}+\rm{H.c.}\non\\&&+\mu\sum_{j}\cos(2\pi\beta j+\phi)n_{j}.
\eea
We further assume that $\ket{\psi (t)}$ has the following expansion
\bea \ket{\psi(t)} ~=~ \sum_{j} ~c_{j}(t)
~e^{-i\int_{0}^{t} dt' E_{j}(t')} \ket{j}. \label{psi1} \eea 
Substituting Eq. \eqref{psi1} in Eq.~\eqref{schr1}, we find that
\beq \frac{dc_{j}}{dt} ~=~ -i ~\sum_{j'} ~\bra{j} V\ket{j'} ~e^{i\int_{0}^{t} 
dt' ( E_{j}(t')-E_{j'}(t'))} c_{j'}(t). \label{cm1} \eeq
One can further recast, $c_{j}=\sum_{n}\,c_{j}^{(n)}$ where $c_{j}^{(n)}$ denotes the coefficient proportional to $V^{n}$. One can now solve Eq. \eqref{cm1} order by order in $V$.  Using the expansion of $c_{j}$ in orders of $V$  mentioned earlier and comparing both sides of Eq. \eqref{cm1}, one can easily show that the zeroth order coefficient, $c_{j}^{(0)}$ satisfies
\bea
\frac{dc_{j}^{(0)}(t)}{dt}&=&0.\label{c0}
\eea
In a similar manner, it can be shown that the first order coefficient, $c_{j}^{(1)}$ satisfies the equation given below
\bea
\frac{dc_{j}^{(1)}(t)}{dt}&=&-i\sum_{j'}\bra{j}V\ket{j'}~e^{i\int_{0}^{t} 
dt' ( E_{j}(t')-E_{j'}(t'))} c_{j'}^{(0)}(t).\non\\\label{c1}
\eea
One can now integrate both sides of Eq. \eqref{c0} to show that $c_{j}^{(0)}(t)=c_{j}^{(0)}(0)$. Moreover, substituting $c_{j}^{(0)}(t)=c_{j}^{(0)}(0)$ in \eqref{c1}, one obtains the following:
\bea
\frac{dc_{j}^{(1)}(t)}{dt}&=&-i\sum_{j'}\bra{j}V\ket{j'}~e^{i\int_{0}^{t} 
dt' ( E_{j}(t')-E_{j'}(t'))} c_{j'}^{(0)}(0).\non\\\label{c11}
\eea
Integrating both sides of Eq. \eqref{c11}, $c_{j}^{(1)}(t)$ turns out to be
\bea
c_{j}^{(1)}(t)&=&c_{j}^{(1)}(0)-i\sum_{j'}\bra{j}V\ket{j'}\int_{0}^{t}\,dt'\times\non\\
&&e^{i\int_{0}^{t'} 
dt' ( E_{j}(t'')-E_{j'}(t''))} c_{j'}^{(0)}(0).
\eea
Without any loss of generality, we further set $c_{j}^{(1)}(0)$ equal to be zero. Therefore, $c_{j}$ to first order in V after one stroboscopic time evolution reduces to
\bea
c_{j}(T)&=&c_{j}^{(0)}(T)+c_{j}^{(1)}(T),\non\\
&=&c_{j}^{(0)}(0)-i\sum_{j'}\bra{j}V\ket{j'}\int_{0}^{T}\,dt'\,e^{i\int_{0}^{t} 
dt' ( E_{j}(t')-E_{j'}(t'))} c_{j'}^{(0)}(0),\non\\
&=&\left(\delta_{jj'}-i\sum_{j'}\bra{j}V\ket{j'}\int_{0}^{T}\,dt'\,e^{i\int_{0}^{t} 
dt' ( E_{j}(t')-E_{j'}(t'))}\right)\times\non\\
&&c_{j'}^{(0)}(0).\non\\ \label{cm2} \eea 
Noting the fact, due to the Floquet degenerate perturbation theory~\cite{Sen_2021}, $c_{j}(0)=c_{j}^{(0)}(0)$, Eq. \eqref{cm2} can be further recast as 
\bea c_{j}(T)& ~=~ &\sum_{j'}\left(I-iH_{F}^{(1)}T\right)_{jj'}c_{j'}(0), \non\\
\bra{j}H_{F}^{(1)}\ket{j'}&~=~&\frac{\bra{j}V\ket{j'}}{T}\,\int_{0}^{T}\,dt\,e^{i\int_{0}^{t} dt' ( E_{j}(t')-E_{j'}(t'))},\non\\
\label{cm3} \eea 
where $I$ denotes the identity matrix and
$H_F^{(1)}$ is the Floquet Hamiltonian to first order in $V$. The non-zero matrix elements of the effective Hamiltonian $H_{F}$ are as follows:
\begin{widetext}
\bea
\bra{j}H^{(1)}_{F}\ket{j+1}&=&\left(1+\lambda \cos(2\pi\beta (j+1/2)+\phi)\right) P_{j},\non\\
\text{where}~~P_{j}&=&e^{i\gamma \sin(\pi\beta(2j+1)+\phi)}~\frac{\sin(\gamma\sin(\pi\beta(2j+1)+\phi)}{\gamma \sin(\pi\beta(2j+1)+\phi)},\non\\
\bra{j}H^{(1)}_{F}\ket{j}&=&\mu\cos(2\pi\beta j+\phi),\non\\
\eea
\end{widetext}
where $\gamma=\frac{\delta\mu T}{2}\sin(\pi\beta)$. Hence, the first order effective Hamiltonian obtained from the first order FPT is given by
\begin{widetext}
\bea
H_{F}^{(1)}&=&\sum_{j}\left(P_{j}\times\left(J+\lambda\cos(2\pi\beta(j+1/2)+\phi)\right)c_{j}^{\dagger}c_{j+1}\non+{\rm H.c.} \right)+\mu \sum_{j}\cos(2\pi\beta j+\phi)n_{j}.
\eea
\end{widetext}
One can further show that the second order effective FPT Hamiltonian, $H_{F}^{(2)}$ vanishes since $H_{F}$ can only have terms which are odd powers in $J,~\lambda$, and $\mu$ due to the symmetry property discussed in Eq. \eqref{sym}~\cite{sreemayee22mobility}. Consequently, the correction next to first order will be third order, and hence, the first-order effective FPT Hamiltonian is a good approximation to the exact effective Floquet Hamiltonian.
\begin{figure}[h!]
\stackon{\includegraphics[width=0.495\columnwidth,height=3.7cm]{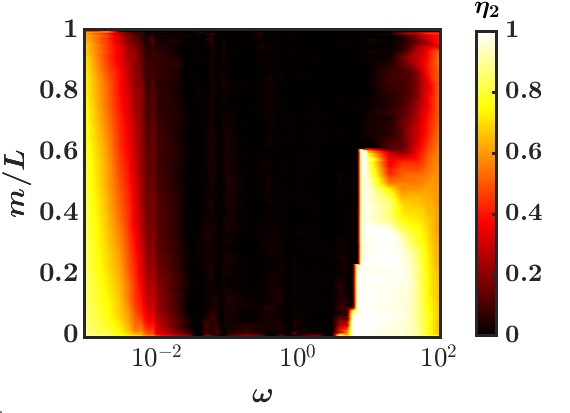}}{(a)}
\stackon{\includegraphics[width=0.495\columnwidth,height=3.7cm]{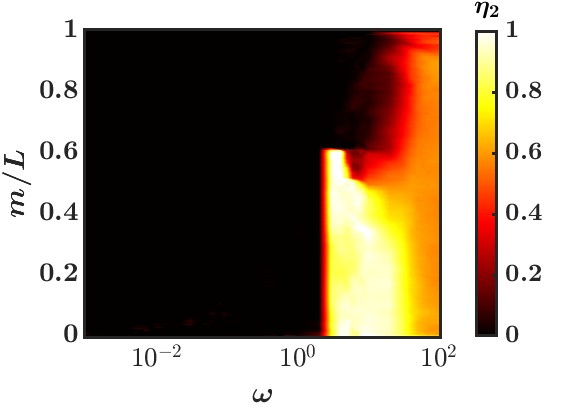}}{(b)}
\caption{{\bf Fractal dimension of Floquet eigenstates obtained from exact numerics and effective FPT Hamiltonian:} (a-b) The variation of $\eta_{2}$ as a function of fractional single-particle Floquet eigenstate index, $m/L$ and driving frequency,  $\om$ for $J=1,~\lambda=0.5$, $\mu=2$, and $\delta\mu=4$. Fig. (a) is obtained from the exact numerical Floquet analysis while Fig. (b) is found from the first order effective FPT Hamiltonian. The dependence of $\eta_{2}$ as a function of $m/L$ and $\omega$ for the first analysis qualitatively agrees with the second one when $10^{-2}<\om<10^2$. However, the effective FPT Hamiltonian fails to capture the delocalization phase when $\om<10^{-2}$  since the effective Hamiltonian is not uniquely defined at such a low frequency regime due to the branch cut ambiguities. } \label{fig11} \end{figure}

In Fig.~\ref{fig11}, we compare the phase diagram of single-particle spectrum-resolved Floquet eigenstates as  a function of $\om$ obtained from the exact numerical Floquet analysis with the one acquired from the first-order FPT Hamiltonian for $J=1,~\lambda=0.5,~\mu=2$ and $\delta\mu=4$. In Figs. \ref{fig11} (a)-\ref{fig11} (b), we see the variation of $\eta_{2}$ (fractal dimension extracted from the scaling of IPR, $I_{m}^{(2)}$, which are sorted in the increasing order of IPR-values for $L=500,~1000$ and $2000$) as a function of $\om$ obtained from the exact numerical Floquet analysis and the first order  FPT Hamiltonian, respectively. Both of the results qualitatively agree with each other when $\om>10^{-2}$. Moreover, both of them in the extremely high frequency regime exhibit critical behavior (similar to the static case as expected), and then slowly flows towards the localized phase with decreasing $\om$-values. However, when $\om<10^{-2}$, $\eta_{2}$ phase diagram obtained from the exact numerical Floquet analysis shows almost delocalized behavior, whereas the first-order FPT Hamiltonian fails to capture this particular regime. The reason for this discrepancy is due to the extremely slow driving limit, which forbids to define the effective FPT Hamiltonian, $H_{F,~FPT}$ in an unambiguous manner due to the following reason. The effective Floquet 
Hamiltonian, $H_F$ in this limit cannot be uniquely defined since 
the eigenvalues $e^{-i \ta_m}$ of the Floquet operator $U$ will not satisfy
$|\ta_m| \ll \pi$ for all the Floquet eigenstates $m$ (also see Sec.~\ref{S8}). Hence $H_F = (i/T) \ln U$ will suffer from branch cut ambiguities. 

Nevertheless, this re-entrant delocalized behavior in extremely slow driving regime observed in the exact numerical Floquet phase diagram is somewhat expected due to the following argument. Note that the driving parameters chosen for this case ($\delta\mu=4$) allows the phase-mixing between the positive $\mu$ and negative $\mu$-direction about the vertical critical line of the static model. To get insight into this phase-mixing, we consider the maximally expanded phase diagram of the static model along the negative $\mu$-axis discussed in Sec.~\ref{S1}. Taking it into account, one can clearly see that periodic driving for this particular set of parameter values causes the delocalized and localized phases to intermix with one another. As discussed in the main text, this kind of phase mixing  initially impels the system toward the localized behavior before finally reaching drive-induced delocalized phase, which we exactly see in Fig. \ref{fig11} (a).
\begin{figure}[h!]
\includegraphics[width=\columnwidth,height=5.8cm]{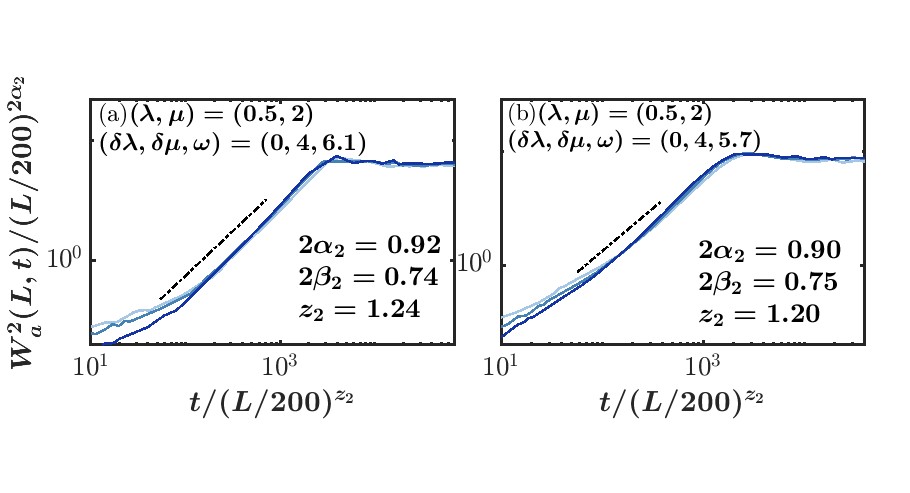}
\caption{{\bf Surface roughness dynamics obtained from exact numerical Floquet analysis and first order FPT Hamiltonian:} (a-b) The stroboscopic time evolution of the square of the approximate surface roughness operator, $W^{2}(L,~t)$ vs time obtained from the exact numerical Floquet calculation and first order FPT Hamiltonian, respectively. In both cases, we consider $J=1,~\lambda=0.5,~\mu=2.0$, $d\mu=4$, and $L=400,~600,~800$. In the first case, we consider $\om=6.1$, and for the second one, we consider $\om=5.7$. The best fitted FV scaling exponents ($2\alpha_2,~2\beta_2,~z_2$) for these cases are as follows: Fig. (a): $(0.92, 0.74, 1.24)$, Fig. (b): $(0.90, 0.75, 1.20)$. } \label{fig13} \end{figure}

\begin{figure*}
\stackon{\includegraphics[width=0.33\linewidth]{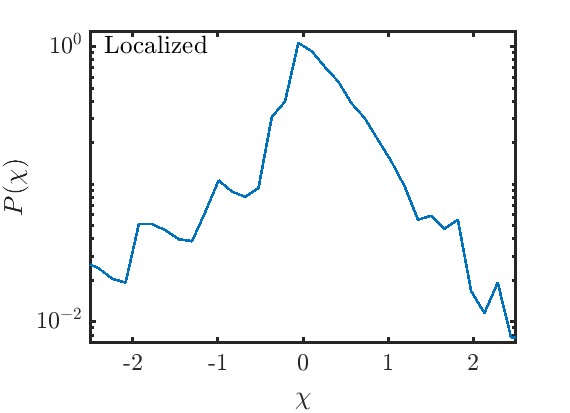}}{(a)}
\stackon{\includegraphics[width=0.33\linewidth]{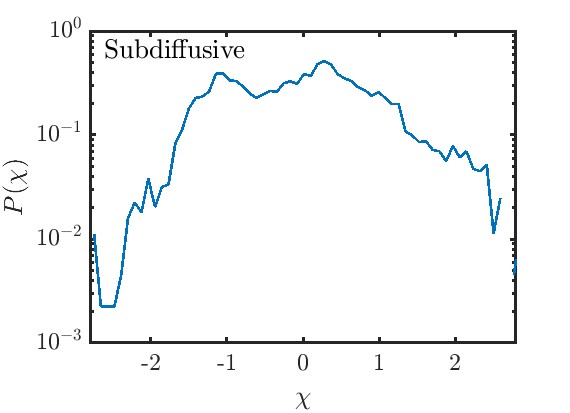}}{(b)}
\stackon{\includegraphics[width=0.33\linewidth]{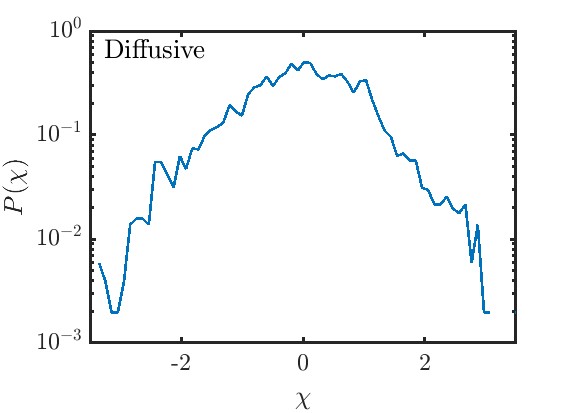}}{(c)}
\stackon{\includegraphics[width=0.33\linewidth]{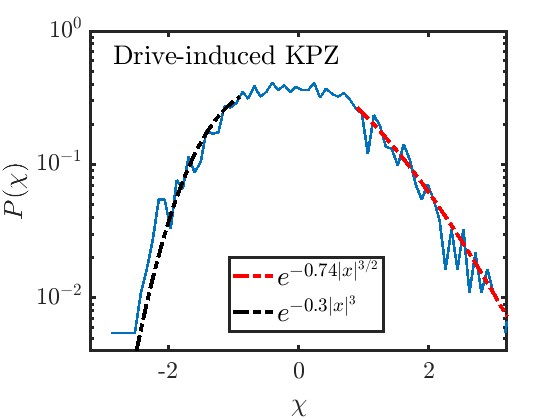}}{(d)}\stackon{\includegraphics[width=0.33\linewidth]{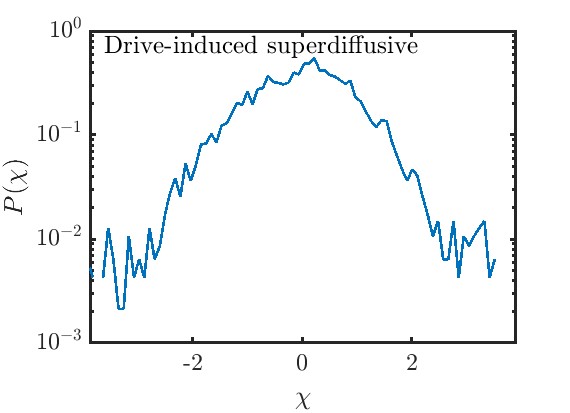}}{(e)}
\stackon{\includegraphics[width=0.33\linewidth]{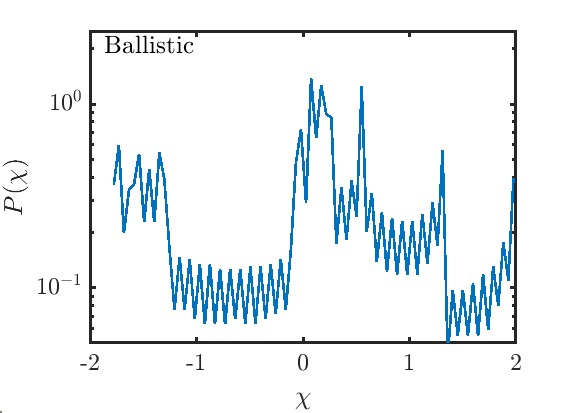}}{(f)}
\caption{{\bf Rescaled height distribution for different dynamical phases:} (a-f) Rescaled surface-height distribution $P(\chi)$ obtained for different dynamical phases for $L=400$ and for 5000 disorder realizations. We consider many-body DW state as the initial state to obtain these plots. The parameter values considered for these cases are (a): $\lambda=0,\mu=3,t=20$ (localized case), (b): $\lambda=1,\mu=1.5, t=1000$ (subdiffusive dynamics), (c): $\lambda=0.5,\mu=2,t=500$ (diffusive dynamics), (d): $\lambda=0.5,\mu=2,\delta\mu=1.5,\om=5,t=200$ (drive-induced KPZ behavior), (e): $\lambda=0.5,\mu=2,\delta\mu=4,\om=5.7$, and $t=500$ (Drive-induced non-KPZ superdiffusive), (f): $\lambda=0,\mu=1$, $t=100$ (ballistic growth dynamics). Moreover, in Fig. (d), we fit the left and right tails of $P(\chi)$ to show the agreement of the asymptotic behavior ($\chi\pm\infty$) with the Tracy-Widom distribution.}
\label{fig14} 
\end{figure*}
In Figs. \ref{fig13} (a)-\ref{fig13} (b), we also compare the square of the approximate surface roughness dynamics stroboscopically in time obtained from the exact numerical Floquet analysis with the one procured from the first order FPT Hamiltonian, respectively. For both cases, we have taken $J=1,~\lambda=0.5,~\mu=2$, and $\delta\mu=4$ (the vertical critical line of the static model subjected to periodic driving along the horizontal direction with large $\delta\mu$) for $L=400,~600,~800$. As shown in Fig. \ref{fig13}, the best fitted FV scaling exponents obtained from the exact numerical Floquet analysis for $\om=6.1$, $(2\alpha_2,2\beta_2,~z_2)=(0.92,~0.74,~1.24)$ are quite close to those from the first order FPT Hamiltonian for $\om=5.7$ with $(2\alpha_2,~2\beta_2,~z_2)=(0.90,~0.75,~1.20)$. This essentially reflects the efficiency of FPT analysis capturing the effective dynamics in this regime. 

\section{Rescaled height distribution for different dynamical phases}
\label{S8}
In this section, we study the
rescaled surface height distribution for different dynamical universality classes appearing in our model in the presence and absence of periodic driving. 
The rescaled height operator for this analysis is defined as follows:
\bea
\chi&=&\frac{h_{L/2}-\langle h_{L/2}\rangle}{\sigma_{h_{L/2}}},
\eea
where $h_{L/2}$ stands for the values of quantum surface height for site $j=L/2$ (defined in the main text) for different disorder realizations. $\langle h_{L/2} \rangle$ and $\sigma_{h_{L/2}}$ denote the mean and standard deviation, respectively of the distribution of $h_{L/2}$. 
In Fig. \ref{fig14}, we examine the rescaled quantum surface-height distribution $P(\chi)$, with zero mean and unit standard deviation, for different dynamical phases appeared in our quantum model. In Fig. \ref{fig14} (a-f), $P(\chi)$ is shown for the localized, subdiffusive, diffusive, drive-induced KPZ, drive-induced superdiffusive (non-KPZ) and ballistic dynamical classes, respectively, with 
the parameters suitably chosen as discussed in the caption of Fig.~\ref{fig14}. 
Interestingly, $P(\chi)$ for each dynamical class appears to be quite distinct from one another, as can be seen from Fig. \ref{fig14}. In the localized case (Fig. \ref{fig14} (a)), the distribution $P(\chi)$ is quite narrow with peak appearing around $\chi\approx0$ whereas in the subdiffusive class (Fig. \ref{fig14} (b)) $P(\chi)$ is broadened and bi-modal with peaks appearing away from $\chi=0$. In the diffusive phase (Fig. \ref{fig14} (c)) a single-peaked broad distribution is observed. Similar broadened distributions are shown in Fig. \ref{fig14} (d) and Fig. \ref{fig14} (e) for the drive-induced KPZ and non-KPZ superfiffusive phases, respectively, although they have their own characteristics which we will discuss later in this section. In the ballistic class (Fig. \ref{fig14} (f)), $P(\chi)$ is not very well-behaved and does not seem to have well-defined asymptotic limits at $\chi\rightarrow\pm \infty$ unlike in the other phases shown here. We note that these distributions, at the moment, are mere numerical observations and may not be the universal ones. However, this demands more of similar (and more careful) studies on various models to come to a conclusion in the future.

Furthermore, in Fig \ref{fig14} (d), we compare the asymptotic behaviors of $P(\chi)$ for the drive-induced quantum KPZ case with the Tracy-Widom (TW) distribution function. The fitting analysis shows that the asymptotic behavior for both $\chi\pm\infty$ follows the TW asymptotes with some fluctuations. Moreover, the drive-induced KPZ distribution appears to be quite distinct from the drive-induced non-KPZ superdiffusive case, which seems to have a more symmetric distribution as shown in Fig. \ref{fig14} (e). It is already known from the literature that the asymptotic behaviors of the rescaled height distribution in the case of classical KPZ universality class follow the TW distribution of the Gaussian unitary ensembles (GUE) for a particular choice of initial condition~\cite{Takeuchi2011}. Recently, this agreement has been also found in the density fluctuations of single-particle wave packet dynamics in the case of two-dimensional Anderson localization~\cite{Senmu2023}. This very similar asymptotic behavior for our drive-induced quantum KPZ case for a many-body (DW type) initial state, therefore, opens up new possibilities for further explorations in the future.
\section{Discussion and conclusion}
\label{conclusion}
In summary, we present an intricate dynamical phase diagram of a one-dimensional fermionic model with correlated disorder, which includes showing the existence of the diffusive and subdiffusive dynamical universality classes of surface roughness with anomalous FV scaling exponents, especially in the critical phase, which is hitherto not reported in the literature due to the rare occurrences of critical phases in the familiar models. Interestingly, the static model has been recently realized experimentally on a superconducting quantum processor with tunable couplers~\cite{li2023observation} and also using ultra-cold atoms~\cite{XIAO20212175}, which can be used as test beds for verifying our results. We also observe that driving this model with periodic square pulse protocol can induce emergent KPZ-like superdiffusive dynamics for certain choices of driving parameters. This is the first observation of this special type of superdiffusive growth dynamics in a periodically driven free fermionic model.
However, this emergent KPZ-like behavior, apparently, can not be explained by taking the semi-classical limit due to the quadratic nature of the (driven) model~\cite{jin2020stochastic,KPZ2019SU2,Ilievski2018,KPZ2022Integrable,keenan2022evidence} and hence opens up scope for further understanding. One can argue that this behavior is intrinsically quantum in nature and arises due to the intricate interplay between periodic driving and the quantum nature of the system (Many-body fermionic states whose evolution governed by the Schrodinger equation), which essentially controls the growth in dynamics, and therefore, has no classical counterpart. 
\par Our study reveals that the interplay between periodic driving and correlated disorder can give rise to new dynamical universality classes~\cite{lezma2019} of surface roughness, which seem to be different from the known ones, and therefore, require further understanding. For the future purpose, it would be worthwhile to examine the surface roughness dynamics in the presence of interaction, 
which might open up new possibilities to unravel the surface growth physics of the many-body localization~\cite{BASKO20061126,vosk,huse2013,Oganesyan2007,pal2010,AAH2015} and interacting Floquet quantum matter~\cite{ponte2015,bordia2017,sierant2023,decker2020,Khemani2016,Else2016, Zhang2017}.

\section*{Acknowledgments}
We would like to thank Diptiman Sen for insightful discussions and for valuable comments on the manuscript. We thank Sumilan Banerjee for fruitful discussions. S.A. thanks MHRD, India for financial support through a PMRF. NR acknowledges support from the Nanyang Technological University and Indian Institute of Science, where the work has been started.

\appendix

\begin{center}
{\bf APPENDIX}    
\end{center}
\section{Symmetries of the static model}
\label{S1}
We begin by discussing the symmetries of the static model (where $\lambda>0$ and $\mu>0$) as considered in the main text, which will allow us to extend the phase diagram for $\lambda<0$ and $\mu<0$. Moreover, this symmetry analysis will be useful later in the discussion to understand the drive-induced phase mixing in the large driving amplitude limit (see Sec.~\ref{S7}). The symmetries of the static model are discussed below.
\begin{enumerate}
\item It should be noted from the phase diagram shown in Fig.~1(a) of the main text that all the states are delocalized for ($0<\lambda<1$, $0<\mu<2$), localized for ($\mu>2$, $\mu>2\lambda$), and critical for ($\lambda>1$, $\mu<2\lambda$). One can further argue that the spectrum will be symmetric about $E=0$ by the argument given below. Followed by the particle-hole transformation, i.e., $c_{j}~\rightarrow~c_{j}^{\dagger}$ and $c_{j}^{\dagger}~\rightarrow~c_{j}$, the static Hamiltonian shown in Eq. 1 of the main text (without loss of generality, we set $\phi$=0 for this analysis) takes the following form:
\bea
H&=&\sum_{j}\left(J+\lambda\cos(2\pi\beta (j+1/2))\right)(c_{j}c_{j+1}^{\dagger}+c_{j+1}c_{j}^{\dagger})\non\\+&&\mu\sum_{j}\cos(2\pi\beta j)~c_{j}c_{j}^{\dagger}, \non\\
&=&-H+\mu\sum_{j}\cos(2\pi\beta j).\label{PH}
\eea
    Now, for sufficiently large system sizes, $\sum_{j}\cos(2\pi\beta j)$ necessarily goes to zero, which, therefore, suggests that the energy spectrum will be symmetric about $E=0$.
     Moreover, this transformation is equivalent to the following changes of the parameters of the static model, $J~\rightarrow~-J$, $\lambda~\rightarrow~-\lambda$, and $\mu~\rightarrow~-\mu$. 
\item Next, one can perform the following symmetry operation on the annihilation operator, namely $c_{j}~\rightarrow~(-1)^{j}c_{j}$, which is necessarily equivalent to $J~\rightarrow~-J$, $\lambda~\rightarrow~-\lambda$, and $\mu~\rightarrow~\mu$. This symmetry transformation along with the previous one imply $J~\rightarrow~J$, $\lambda~\rightarrow~\lambda$, and $\mu~\rightarrow~-\mu$. 
\item There is one more symmetry transformation, which one can perform using the property of an irrational number. First, one should note that both quasiperiodic hopping and quasiperiodic potential contain $\cos(2\pi\beta j)$ kind of terms, where $\beta$ is the golden ratio. Under the rational approximation, $\beta$ is equal to $\frac{P}{L}$, where $L$ is the system size and $P$ is the integer closest to $(\frac{\sqrt{5}-1}{2})L$. Using this fact, one can further perform the following transformation, namely, $j~\rightarrow~j+\frac{L}{2P}$, followed by which, both quasiperiodic potential and quasiperiodic hopping change signs. Therefore, this transformation suggests $J~\rightarrow~J$,  $\mu~\rightarrow~-\mu$, and $\lambda~\rightarrow~-\lambda$.
\end{enumerate}
Taking these three transformations into account, one can now maximally expand the phase diagram along the negative $\lambda$ and negative $\mu$ direction of the $\lambda-\mu$ parameter plane. The extension of the phase diagram along the negative axes indicates the following:
\begin{enumerate}
    \item Taking the second symmetry into account, the phase diagram in the second quadrant of the $\lambda-\mu$ plane tells us: there will be a delocalized regime for ($0<\lambda<1$, $-2<\mu<0$), a localized phase for ($\mu<-2,~-2\lambda$), and a critical phase for ($2\lambda>2,~-\mu$).
    \item Taking the third symmetry into consideration, the phase diagram can be extended in the third quadrant of the $\lambda-\mu$ plane: a delocalized regime for ($-1<\lambda<0$, $-2<\mu<0$), a localized regime for ($\mu<-2,~2\lambda$) and a critical regime for ($2\lambda<-2,~\mu$).
    \item Including all three symmetry transformation discussed above implies that phase diagram will remain unaltered for $J~\rightarrow~J$, $\lambda~\rightarrow~-\lambda$ and $\mu~\rightarrow~\mu$. With this transformation, one can further extend the phase diagram in the fourth quadrant of the $\lambda-\mu$ parameter plane. In this case, the phase diagram will be the following: a delocalized regime for ($-1<\lambda<0$, $0<\mu<2$), a localized regime for ($\mu>2,~-2\lambda$), and a critical regime for ($2\lambda<-2,~-\mu$).
\end{enumerate}

\section{Finite-size scaling of the steady-state entanglement entropy and phase transitions in the static model}
\label{S2}
In this section, we will discuss the finite-size scaling analysis of the entanglement entropy ($S_\infty$) of the nonequilibrium steady-state on the critical phase boundaries, i.e., the AAH point, the triple point, and three critical lines. To perform the appropriate scaling analysis, we normalize $S_\infty$ by the thermal page value $[S_{T}=(L/2)\ln 2 - 0.5]$, which is expected to be scale invariant at criticality. Therefore, the finite-size scaling ansatz for the entanglement entropy is as follows~\cite{mbc_phase}
\bea
\frac{S_{\infty}}{S_{T}}=f(\delta\tau\,L^{1/\nu}).\label{fin2}
\eea 
where $S_{\infty}$ is the saturation value of the entanglement entropy after being approached the long time limit, $L$ stands for the system size. the correlation length $\xi$ diverges at the critical phase boundaries as $\xi~\sim~\delta\tau^{-{\nu}}$, where $\delta\tau=|\tau-\tau_c|$ is the distance from the critical value in the parameter space, and ${\nu}$ denotes the critical exponent associated with the correlation length. 
\begin{figure*}
\subfigure[]{\includegraphics[width=0.33\linewidth]{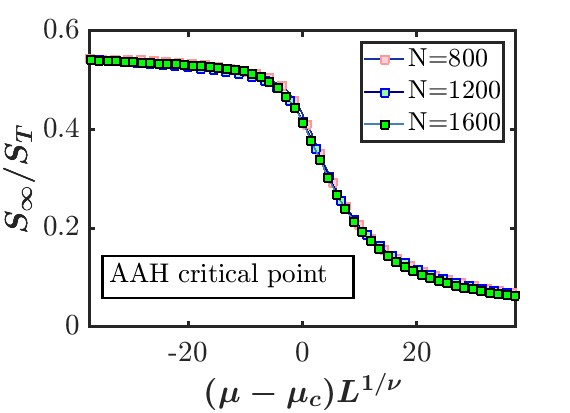}}%
\subfigure[]{\includegraphics[width=0.33\linewidth]{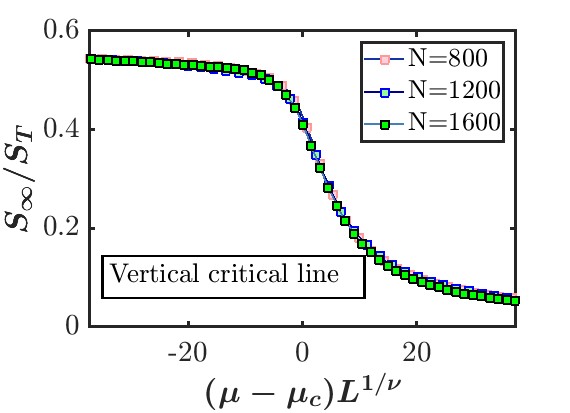}}%
\subfigure[]{\includegraphics[width=0.33\linewidth]{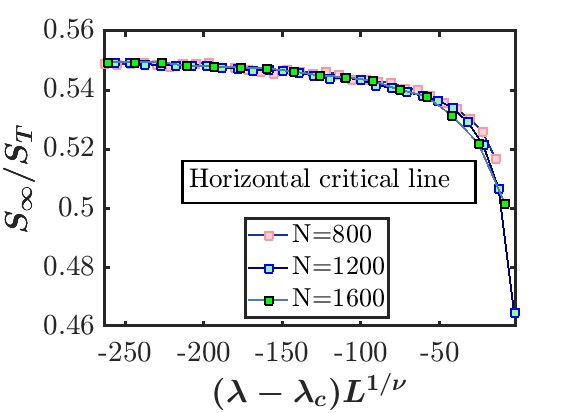}}\\
\subfigure[]{\includegraphics[width=0.33\linewidth]{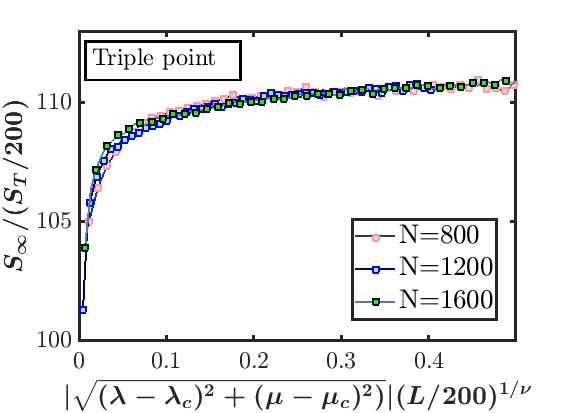}}%
\subfigure[]{\includegraphics[width=0.33\linewidth]{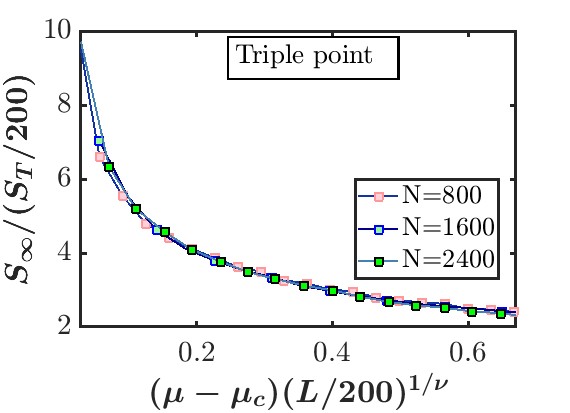}}%
\subfigure[]{\includegraphics[width=0.33\linewidth]{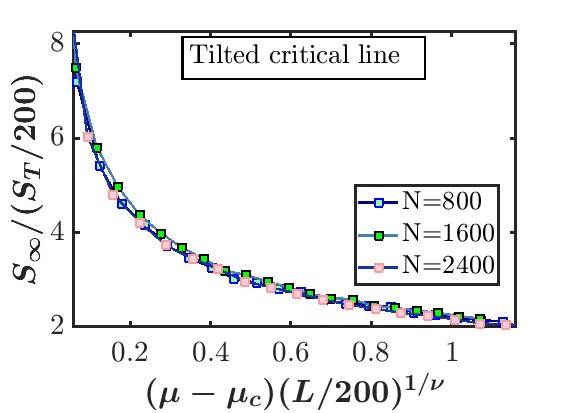}}\\
\label{entsca}
\caption{{\bf Finite-size scaling of the saturation value of entanglement entropy $S_{\infty}/S_{T}$ }:
Fig. (a): Scaling collapse for transition around the AAH point $(\lambda=0$,~$\mu=2)$, Fig. (b): similar analysis around a point ($\lambda=0.5$,~$\mu=2$) on the vertical critical line, Fig. (c): for a point $(\lambda=1$,~$\mu=1.5)$ on the horizontal critical point while approaching from the delocalized side of the phase diagram, Fig. (d-e): same analysis for the triple point with $\lambda=1$ and $\mu=2$, while  the critical line being approached from the delocalized and localized sides of the phase diagram, respectively. Fig. (f): similar analysis for a point $(\lambda=1.5$, $\mu=3)$ on the tilted critical line with  with the point being approached from the localized side of the phase diagram. Both for Figs. (a-b), we see the best scaling collapse for $\nu=1$. 
For Fig. (c), we see the best scaling collapse for ${\nu}=0.64$. For Fig.(d-e), the estimated values of ${\nu}$ are found to be $\approx~0.6$ and $0.44$, respectively, while being approached from the delocalized and localized sides of the phase diagram. Fig. (f): We observe the best scaling collapse for ${\nu}=0.45$.}
\label{entanglesca}
\end{figure*}
As per Eq. \eqref{fin2}, one would expect that $S_{\infty}/S_{T}$ vs $\delta\tau L^{1/\nu}$ to exhibit a scale collapse to a single curve according to Eq. \eqref{fin2}.

As shown in Fig. \ref{entanglesca} (a)-\ref{entanglesca} (b), we first analyze the finite-size scaling of the entanglement entropy for the AAH critical point and the vertical critical line ($\lambda=0.5,~\mu=2$) of the static model, separating the extended and localized phases. For both cases, we see the best scaling collapse for $\nu=1$, which agrees with the earlier findings~\cite{KZ_aah}. In Fig. \ref{entanglesca} (c), we repeat the same analysis for the horizontal critical line with $\lambda=1,~\mu=1.5$, which separates the extended phase from the critical one. For this case, we observe $\nu~\approx~0.64$ when being approached from the delocalized side of the critical line. In Fig. \ref{entanglesca} (d)- \ref{entanglesca} (e), we do the similar analysis for the triple point with $\lambda=1$ and $\mu=2$ when being approached from the delocalized and localized sides of phase diagram, respectively. For these two cases, we see the best scaling collapse for $\nu\approx0.6$ and $\nu\approx0.44$, respectively. In Fig. \ref{entanglesca} (f), we perform the same analysis for the tilted critical line with $\lambda=1.5$ and $\mu=3$ while approaching from the localized side of the critical line, and the corresponding value of $\nu\approx~0.45$. Interestingly, the triple point shows two different values of the critical exponent $\nu$ when approached from the delocalized phase and localized phase, respectively. The new kind of criticality of the quantum triple point~\cite{belitz2017quantum} demands a more careful investigation, which goes beyond the scope of current work. Scaling collapse in the critical phase is not possible due to strong fluctuations in the data, which presumably originates due to characteristic huge quantum fluctuations in the critical phase. 
\section{The growth of the exact surface roughness operator at the AAH critical point and inside the critical phase}
\begin{figure}[h!]
\includegraphics[width=\columnwidth,height=5.5cm]{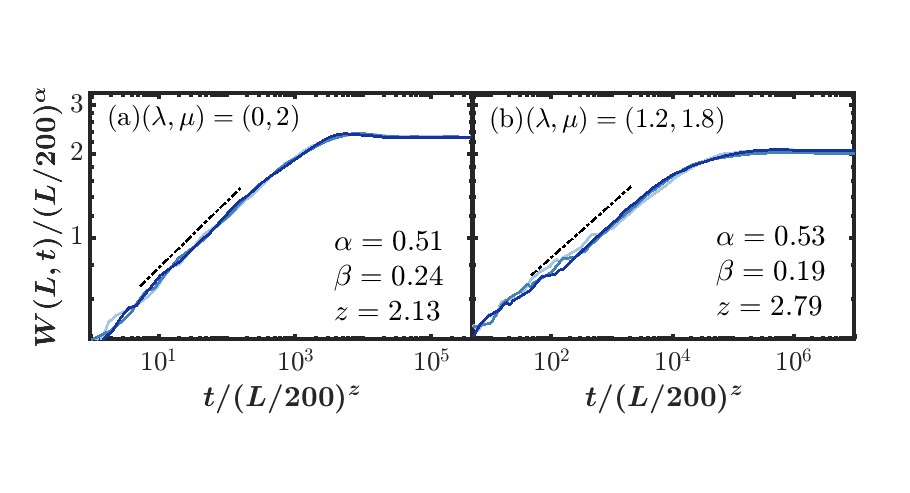}%
\caption{{\bf The dynamics of the surface roughness operator at the AAH critical point and inside the critical phase}: (a-b) The variation of the exact surface roughness operator $W(L,t)$ with time for $L=400$, $600$ and $800$. The parameter values chosen for these plots are as follows: Fig. (a): $\lambda=0,~\mu=2$ (AAH critical point), Fig. (b): $\lambda=1.2,~\mu=1.8$ (a point inside the critical phase). In the first cases, the dynamics show diffusive growth in time with best fitted FV scaling exponents $(\alpha,\beta,z)$=$(0.51,0.24,2.13)$. In Fig. (b), we see subdiffusive growth in time with the best fitted FV scaling exponents $(\alpha,\beta,z)=(0.53,0.19,2.79)$. }
\label{exactw}
\end{figure}
Here we show the dynamics of the exact surface roughness operator $W(L,t)$ with time for the AAH critical point and for the critical phase. In Fig. \ref{exactw} (a-b), we consider the AAH critical point $(\lambda,\mu)=(0,2)$ of the static model and a point inside the critical phase of the static model with $\lambda=1.2$ and $\mu=1.8$, respectively. In Fig. \ref{exactw} (a), we see that the surface roughness exhibits almost the diffusive dynamical behavior with the best fitted FV scaling exponents $(\alpha,\beta,z)=(0.51,0.24,2.13)$, which agrees with the earlier findings~\cite{thouless1994critical}. On the other hand, the critical phase demonstrates subdiffusive dynamical behavior with the scaling exponents $(\alpha,\beta,z)=(0.53,0.19,2.79)$, as shown in Fig. \ref{exactw}(b).

\section{Fractal dimension of the single-particle Floquet eigenstates of the periodically driven model}
\label{S4}
\begin{figure}[h!]
\subfigure[]{\includegraphics[width=0.53\linewidth,height=3.6cm]{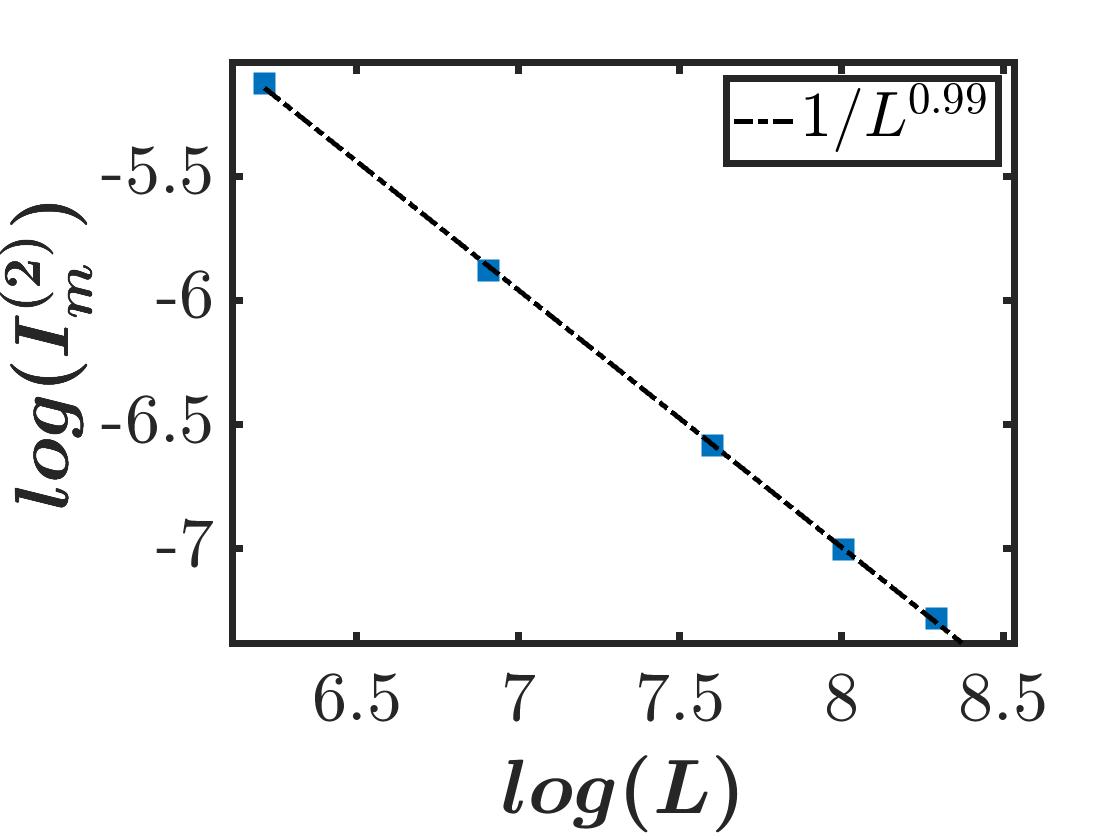}}%
\subfigure[]{\includegraphics[width=0.53\linewidth,height=3.6cm]{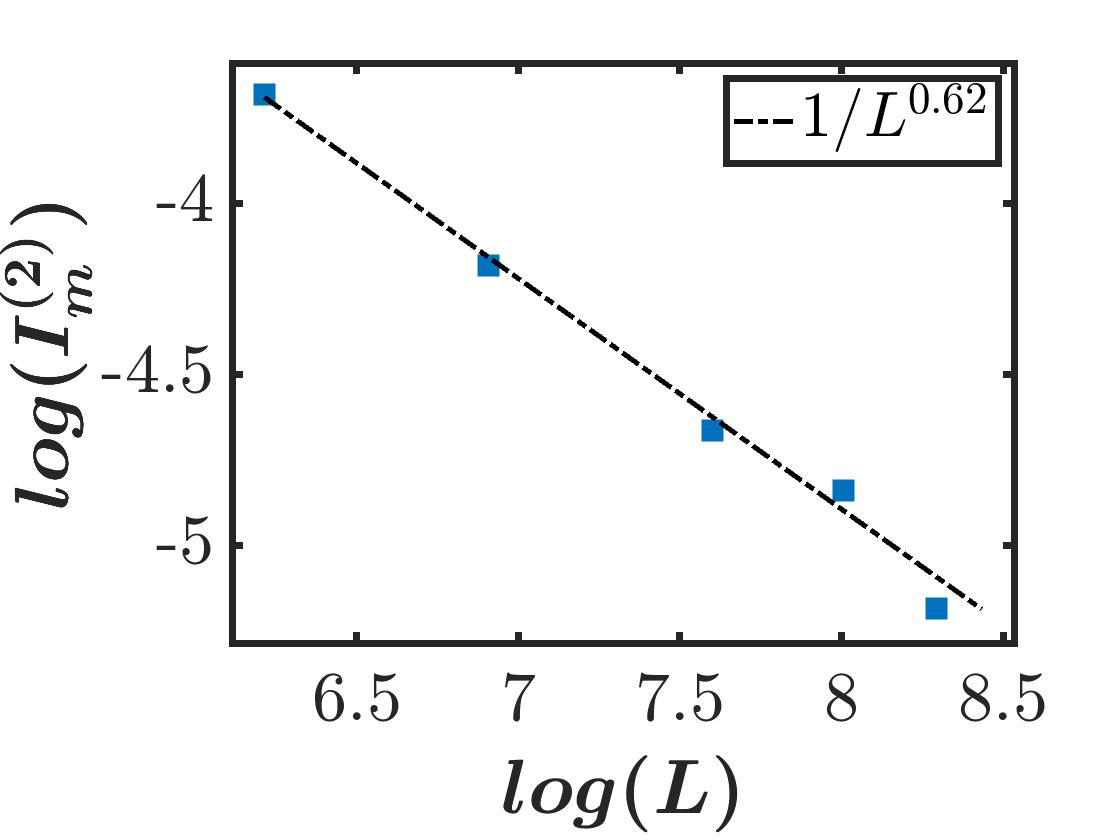}}\\
\subfigure[]{\includegraphics[width=0.53\linewidth,height=3.6cm]{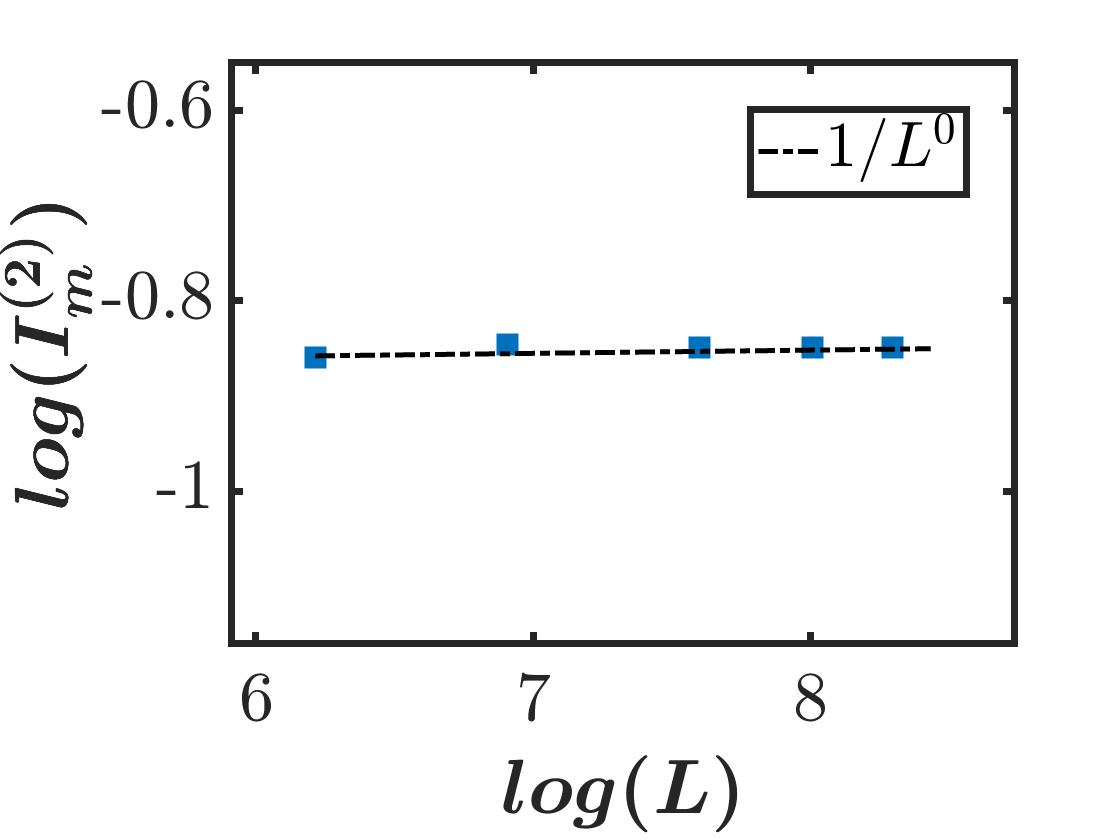}}%
\caption{ {{\bf The scaling exponents of the inverse participation ratios of the single-particle spectrum-resolved Floquet eigenstates}:} (a-c) The inverse participation ratio ($I_{m}^{(2)}$) of the single-particle spectrum-resolved Floquet eigenstates are plotted for three different system sizes, $L=500,~1000$ and $2000$, $3000$ and $4000$ in log-log scale. The scaling exponent ($\tau_{2}$) can be easily extracted from the slope of $log(I_{m}^{2})$ vs $log(L)$ plot by a linear fitting analysis. Fig. (a): In this case, we consider a single-particle Floquet eigenstate with $m/L=0.5$$J=1$, $\lambda=0.5$, $\mu=2$, $\delta\mu=1.5$, and $\om=0.005$. Fig. (b): Here we consider $J=1$, $\lambda=1.5$, $\mu=3$, $\delta\mu=2.5$, and $\om=30$. The same analysis as Fig. (a) is performed for the single-particle Floquet eigenstate with $m/L=0.7$. Fig. (c): For this particular case, we consider $J=1$, $\lambda=1.5$, $\mu=3$, $\delta\mu=2.5$, $\om=0.5$, and the single-particle Floquet eigenstate with $m/L=0.5$. The values of $\tau_{2}$ extracted for these cases obtained from the numerical fitting analysis are $1~ (\text{extended}),~0.62~(\text{critical})$, and $0.00~(\text{localized})$, respectively.}\label{fig1111}
\end{figure}

In this section, we will analyze the extraction procedure of the fractal dimension from the scaling of the inverse participation ratio (IPR) for the single-particle Floquet eigenstates using the numerical fitting procedure. To determine if any of the Floquet eigenstates exhibit multifractal behavior,
we calculate IPR, also defined in the main text as
\beq I_m^{(2)} ~=~ \sum_j ~|\psi_m (j)|^{4}, \label{imp} \eeq
and study its scaling with the system size 
$L$~\cite{caste,evers}. 
If $I_m^{(2)}$ scales as $L^{-\tau_2}$, 
then $\tau_2 = 1$ for extended states (since $|\psi_m (j)|^2$ 
is of the order of $1/L$ for all $j$ for such states), and 
$\tau_2 = 0$ for localized states (since $|\psi_m (j)|^2$
for such states is of order 1 over a finite region whose size
remains constant as $L \to \infty$). Multifractal or critical states 
typically have $0 < \tau_2 < 1$. In Figs. \ref{fig1111} (a-c), we consider three single-particle Floquet eigenstates for three different parameter values, namely, (a) $J=1$, $\lambda=0.5$, $\mu=2$, $\delta\mu=1.5$, and $\om=0.005$ (vertical critical line of the static model subjected to periodic driving in the slow driving limit), (b-c) $J=1$, $\lambda=1.5$, $\mu=3$, $\delta\mu=2.5$, $\om=30$, and $0.5$, respectively (tilted critical line of the static model subjected to periodic driving in the fast and slow driving regime, respectively). We then extract the scaling exponent of IPR ($\tau_2$) by performing a  linear fit of $log(I_{m}^{(2)})$ vs $log(L)$ plot. As shown in Fig. \ref{fig1111} (a-c), the single-particle Floquet eigenstate show extended, (multi)fractal and localized behavior with $\tau_2=0.99$, $0.62$, and $0$, respectively, which have been extracted from the linear fitting analysis. This is how the dynamical phases, as shown in Figure 3 of the main text, has been determined.
\section{
More details on the dynamics of the driven model}
\label{S5}
\subsection{Equivalence between the approximate surface-roughness operator and half-chain entanglement entropy in the case of periodically driven model}
\begin{figure}[h!]
\stackunder{\includegraphics[width=0.495\linewidth,height=3.6cm]{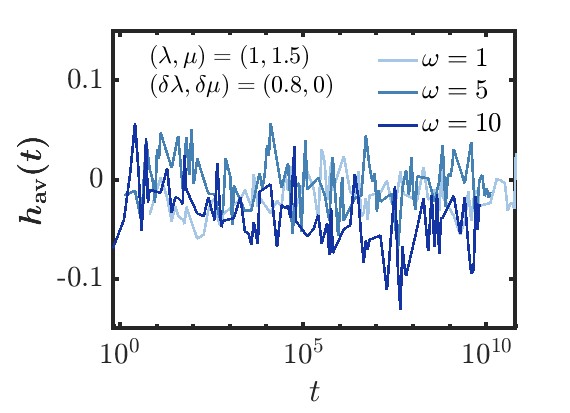}}{(a)}
\stackunder{\includegraphics[width=0.495\linewidth,height=3.6cm]{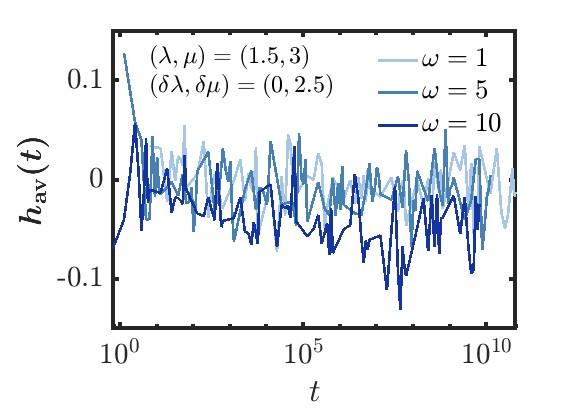}}{(b)}
\stackunder{\includegraphics[width=0.495\linewidth,height=3.6cm]{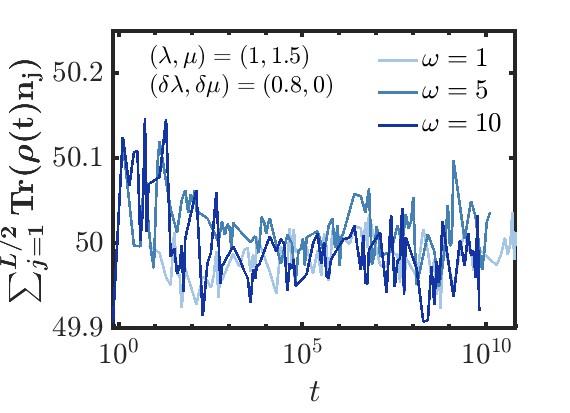}}{(c)}
\stackunder{\includegraphics[width=0.495\linewidth,height=3.6cm]{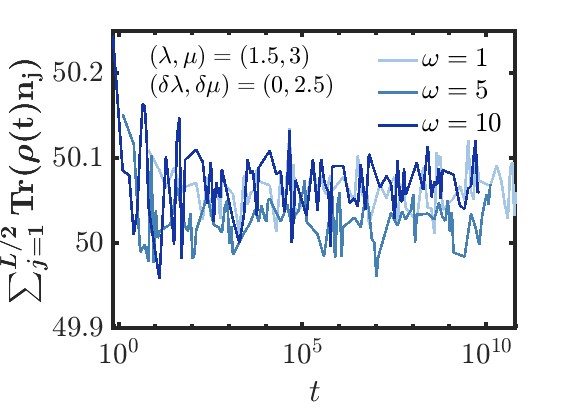}}{(d)}
\caption{{\bf Stroboscopic time evolution of $\mathbf {h_{\rm{av}}(t)}$ and $\mathbf{ \sum_{j=1}^{L/2} {\rm Tr}} \mathbf{(\rho(t)n_{j})}$ of the periodically driven model:} (a-b) The stroboscopic evolution of $h_{\rm{av}}(t)$ vs $t$ for the parameter values, Fig. (a): $\lambda=1,~\mu=1.5,~\delta\lambda=0.8$, Fig. (b): $\lambda=1.5,~\mu=3,~\delta\mu=2.5$. For both cases, we consider slow to fast driving limits with $\om=1,~5,~10$, and for $L=200$. In both cases, $h_{\rm{av}}(t)$ remains close to zero for all time. (c-d) The stroboscopic time evolution of $\sum_{j=1}^{L/2}{\rm Tr}(\rho(t)n_{j})$ are shown with time for the parameter values same as Figs. (a-b). In both cases , $\sum_{j=1}^{L/2}{\rm Tr}(\rho(t)n_{j})=\frac{\nu L}{2}\simeq50$ during the stroboscopic dynamics just as the static model, which is expected since $\nu=1/2$ and $L=200$ for these particular cases. }\label{figs112}
\end{figure}

Just like the static case, the equivalence between approximate surface roughness operator, $W_a^{2}(t)$ and von Neumann entanglement entropy, $S_{L/2}(t)$~\cite{fujimoto2021dynamical} holds even for the periodically driven model in the case of stroboscopic time evolution. To show that, in Figs. \ref{figs112} (a-d), we examine the stroboscopic time evolution of $h_{\rm{av}}(t)$ and $\sum_{j=1}^{L/2}{\rm Tr}(\rho(t)n_{j})$ for the parameter values as follows: Figs. \ref{figs112} (a)~-~\ref{figs112} (c): $\lambda=1,~\mu=1.5,~\delta\lambda=0.8$ (driving the horizontal critical line of the static model along the vertical direction with finite $\delta\lambda$), and Figs. \ref{figs112} (b)~-~\ref{figs112} (d): $\lambda=1.5,~\mu=3,~\delta\mu=2.5$ (driving the tilted critical line of the static model along the horizontal direction with finite $\delta\mu$). In both cases, we consider periodic driving ranging from slow to fast driving limit with $\om=1,~5$ and $10$ as for examples. During the stroboscopic time evolution, $h_{\rm{av}}(t)$, as shown in Figs. \ref{figs112} (a-b), and $\sum_{j=1}^{L/2}{\rm Tr}(\rho(t)n_{j})$, as shown in Figs. \ref{figs112} (c-d), remain close to zero and $\frac{\nu L}{2}$ ($\nu$ being the filling fraction of the many-body initial state) similar to the static case, revealing that stroboscopic Floquet dynamics satisfy the conditions (i) and (iii) as mentioned earlier in the static case. These conditions lead to the equivalence between the square of the approximate surface roughness and von Neumann entanglement entropy even  during the stroboscopic dynamics.

\subsection{Localized states with no well-defined FV scaling exponents}
\begin{figure}[h!]
\subfigure[]{\includegraphics[width=0.53\linewidth]{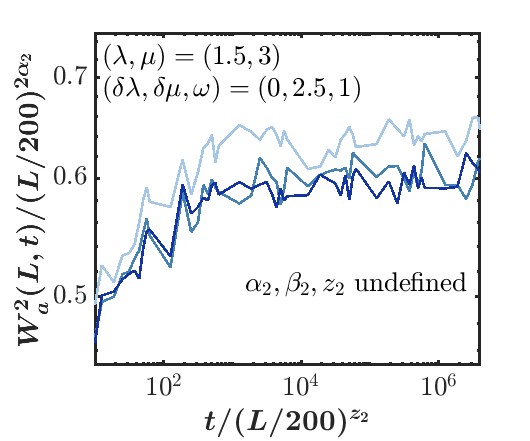}}%
\subfigure[]{\includegraphics[width=0.53\linewidth]{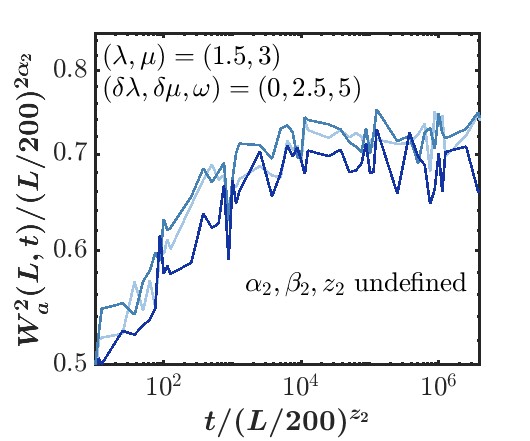}}\\
\subfigure[]{\includegraphics[width=0.53\linewidth]{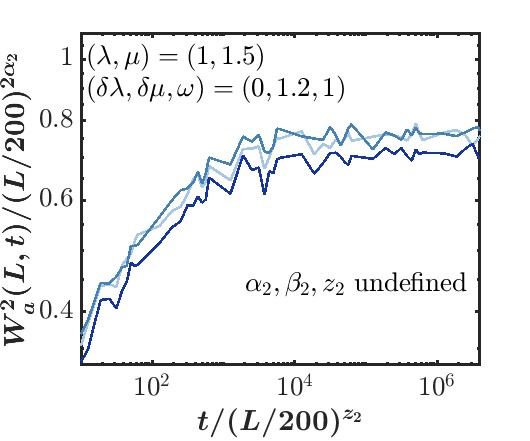}}%
\caption{{\bf Absence of FV scaling in localized phase of the periodically driven model:} (a-b) The stroboscopic variation of approximate $W^{2}(L,t)$ with time for the tilted critical line ($\lambda=1.5,~\mu=3.0$) driven with $\delta\lambda=0$ and $\delta\mu=2.5$ in comparatively slow driving regime ($\om=1$ and $\om=5$, respectively). (c) The same variation for the horizontal critical line ($\lambda=1$, $\mu=1.5$) subjected to periodic driving with $(\delta\lambda,\delta\mu,\om)=(0,1.2,1)$. In all three cases, we see localized behavior with no well-defined FV scaling behavior. } \label{figs4} \end{figure}

Here we will discuss the FV scaling in the case of drive-induced localized states. As shown in Figs. \ref{figs4} (a-b), the tilted critical line of the static model ($\lambda=1.5,~\mu=3$) is subjected to inter-phase driving (critical and localized phases) with $(\delta\lambda,\delta\mu)=(0,2.5)$ in the comparatively slow driving regime ($\om=1$ and $\om=5$). In Fig. \ref{figs4} (c), we drive the horizontal critical line in the identical manner with $(\delta\lambda,\delta\mu,\om)=(0,1.2,1)$. In all three cases, we see no well-defined scaling exponents since the surface roughness saturates to a value that does not scale up with the system size properly. Hence, here FV scaling is absent for localized phases. 
\begin{figure}[h!]
\subfigure[]{\includegraphics[width=0.52\linewidth]{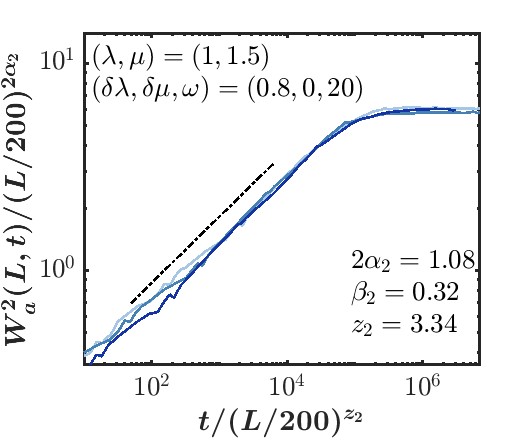}}%
\subfigure[]{\includegraphics[width=0.52\linewidth]{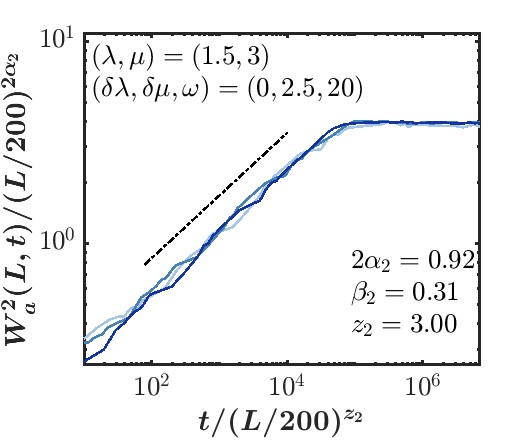}}\\
\subfigure[]{\includegraphics[width=0.52\linewidth]{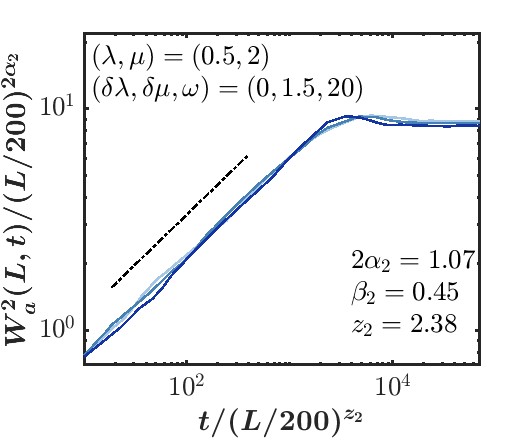}}%
\caption{{\bf Dynamics of driven systems in extremely fast driving regime:} (a-c) The stroboscopic time evolution of approximate $W^{2}(L,t)$ vs time where the horizontal, tilted and vertical critical lines are subjected to periodic driving with Fig. (a): $\lambda=1$, $\mu=1.5$, $\delta\lambda=0.8$, $\delta\mu=0$, and $\om=20$, Fig. (b): $\lambda=1.5$, $\mu=3$, $\delta\lambda=0$, $\delta\mu=2.5$, and $\om=20$, Fig. (c):  $\lambda=0.5$, $\mu=2$, $\delta\lambda=0$, $\delta\mu=1.5$, and $\om=20$, respectively. The best fitted FV scaling exponents, $(2\alpha_2, 2\beta_2, z_2)$ for these cases are found to be: Fig. (a): $(1.08, 0.32, 3.34)$, Fig. (b): $(0.92,~0.31,~3.00)$, Fig. (c) $(1.07,~0.45,~2.38)$, which are quite close to the static case.} \label{figs6}
\end{figure}
\subsection{The surface-roughness dynamics in the extremely fast driving regime}
The dynamics of the periodically driven model in the extremely high frequency regime converge to the static model, which can be seen in Fig. \ref{figs6}. As shown in Figs. \ref{figs6} (a-c), the horizontal ($\lambda=1,~\mu=1.5$), tilted ($\lambda=1.5,~\mu=3$) and vertical critical lines of the static model are subjected to periodic driving in the high-frequency regime with the parameter values, $(\delta\lambda,\delta\mu,\om)=(0.8,0,20)$, $(0,2.5,20)$, and $(0,1.5,20)$, respectively. In first two cases, the square of the approximate surface roughness operator show subdiffusive dynamics with the best fitted FV scaling exponents $(2\alpha_2,~2\beta_2,~z_2)$=$(1.08,~0.32,~3.34)$ and $(0.92,0.31,3.00)$, respectively, while the last one demonstrates almost diffusive dynamics with FV scaling exponents $(2\alpha_2,~2\beta_2,~z)=(1.07,0.45,2.38)$. These behaviors are found to be the same as observed in the static model, which one can anticipate due to reduced amount of drive-induced phase mixing in the high-frequency limit.

\section{Entanglement entropy of the driven model using the correlation matrix method}
\label{S9}
In this section, we will briefly discuss the formulation of the correlation matrix method to analyze the stroboscopic dynamics of a generic non-interacting Floquet system. Before proceeding further, we first briefly discuss how to compute the dynamics of the entanglement entropy of a static model with a non-interacting Hamiltonian $H$ starting from an many-body initial state $\ket{\psi_{\rm{in}}}$, which is not the many-body eigenstate of $H$. To do so, let us first consider a generic non-interacting Hamiltonian given by
\bea
H=\sum_{i\neq j}t_{ij}c_{i}^{\dagger}c_{j}+\rm{H.c.},
\eea
where $c_{i}$ and $c_{i}^{\dagger}$ are the fermionic annihilation and creation operator, respectively. Followed by the diagonalization, such a Hamiltonian takes the following form
$H=\sum_{k=1}^{N}\epsilon_{k}\tilde{c}^{\dagger}_{k}\tilde{c}_{k}$,
where $\tilde{c}_{k}=\sum_{j=1}^{N}\psi_{k}(j)c_{j}$. Setting $\hbar=1$, the time evolution of the annihilation operator can be found using the Heisenberg equation of motion
$\frac{d\tilde{c}_{k}(t)}{dt}=\frac{1}{i}[\tilde{c}_{k},H]=\frac{1}{i}\epsilon_{k}\tilde{c}_{k}$,
which implies $\tilde{c}_{k}=e^{i\epsilon_{k}t}\tilde{c}_{k}(0)$.
To study the dynamics, we consider a many-body density-wave (DW) type of initial state given below
\bea
\ket{\psi_{\rm{in}}}=c_{2}^{\dagger}c_{4}^{\dagger}......c_{N}^{\dagger}\ket{0},
\eea
where $N$ being the system size with an even number of lattice sites, and we further assume that the system is at half-filling, i.e., $N_{p}=N/2$. Next, we will construct the correlation matrix~\cite{IngoPeschel2003,IngoPeschel2012,Peschel_2009,roy2018entcontour} in a sub-system (A) of size $L$ for this initial many-body quantum state, which is, $C_{ij}(t)=\bra{\psi_{\rm{in}}}c_{i}^{\dagger}c_{j}\ket{\psi_{\rm{in}}}$, where $i,~j\in A$. Following the usual procedure~\cite{IngoPeschel2003,IngoPeschel2012,Peschel_2009,roy2018entcontour}, one can easily deduce that the elements of the correlation matrix of the full system takes the following form
\bea
\bra{\psi_{\rm{in}}}c_{i}^{\dagger}c_{j}\ket{\psi_{\rm{in}}}=
\sum_{k,k'=1}^{N}\sum_{i'=2,4,....} && \psi_{i}(k)\psi_{i'}^{\ast}(k')\psi_{i'}(k)\psi_{i'}^{\ast}(k')\non\\&&~\times e^{-i(\epsilon_{k}-\epsilon_{k'})t}
\label{corrm}
\eea
The entanglement entropy $S_{A}(t)$ followed by the diagonalization of a $L\times L$ block of the correlation matrix associated with the sub-system A is given by
\bea
S_{A}(t)=-\sum_{m=1}^{L}\left[\lambda_{m}\ln \lambda_{m}+(1-\lambda_{m})\ln(1-\lambda_{m})\right],\non\\
\eea
where $\lambda_{m}$ stands for the $m$-the eigenvalue of the $L\times L$ block of the correlation matrix. 

Now, we will generalize this procedure for the stroboscopic closed Floquet dynamics. For a generic periodically driven system, one can define an effective time-independent Floquet Hamiltonian, $H_{F}=\frac{i}{T}\ln(U_{F}(T))$ as long as the Floquet eigenvalues, $e^{-i\theta_{m}}$ satisfy the condition, $|\theta_{m}|<<\pi$ for all the Floquet eigenstates, where $T$ and $U_{F}$ being the driving period and the Floquet evolution operator, respectively. Therefore, in this case, the static formalism will hold with a slight modification, i.e., the Hamiltonian of a static model, $H$ will get replaced by the effective Floquet Hamiltonian, $H_{F}$ as far as the stroboscopic Floquet dynamics is concerned. Interestingly, the correlation matrix method holds even if the condition,  $|\theta_{m}|<<\pi$  is violated (mostly in extremely slow driving regime, which is closely connected to the adiabetic limit), and the reason for this is as follows. Note that in the last line of Eq. \ref{corrm}, the eigenvalues of a generic static Hamiltonian appear only in the exponential factors, i.e., $e^{-i\epsilon_{k}t}$ factors, which turns out to be crucial for its generalization to the stroboscopic time evolution of a generic closed non-interacting Floquet quantum system. In this scenario, although $H_{F}$ is not uniquely defined, the Floquet eigenvalues, $e^{-i\epsilon_{m}T}$ (also well-defined for all driving frequencies) still can be defined by diagonalizing the Floquet evolution operator, $U_{F}$ (setting $\hbar=1$),
\bea
U_{F}\ket{m}=e^{-i\epsilon_{m}T}\ket{m},\label{DZ}
\eea
where $\ket{m}$ denotes the $m$-th single-particle Floquet eigenfunction. Substituting the Floquet eigenvalues and single-particle Floquet eigenstates obtained from Eq. \eqref{DZ} in Eq. \eqref{corrm}, one can easily generalize the correlation matrix formalism to analyze the stroboscopic Floquet dynamics even if the effective Floquet Hamiltonian is ambiguous.
\bibliography{refs}
\end{document}